\documentclass[12pt]{iopart}

\pdfoutput=1

\expandafter\let\csname equation*\endcsname\relax
\expandafter\let\csname endequation*\endcsname\relax

\usepackage{graphicx, bm, physics, hyperref, amsmath, amssymb}
\usepackage[normalem]{ulem}
\usepackage[dvipsnames]{xcolor}
\makeatletter

\def\graphicscale{\twocolumn@sw{0.3}{0.4}}
\def\graphicthreescale{\twocolumn@sw{0.3}{0.4}}

\begin{document}

\title{Decoherence and energy flow in the sunburst
  quantum Ising model}

\author{Alessio Franchi, Davide Rossini, and Ettore Vicari}
\address{Dipartimento di Fisica dell'Universit\`a di Pisa
  and INFN, Largo Pontecorvo 3, I-56127 Pisa, Italy}

\date{\today}

\vspace{10pt}

\begin{indented}
\item[]Authors are listed in alphabetic order
\end{indented}

\begin{abstract}
  We study the post-quench unitary dynamics of a quantum sunburst spin
  model, composed of a transverse-field quantum Ising ring which is
  suddenly coupled to a set of independent external qubits along the
  longitudinal direction, in a way to respect a residual translation
  invariance and the Ising $\mathbb{Z}_2$ symmetry.  Starting from the
  different equilibrium quantum phases of the system, we characterize
  the decoherence and the energy storage in the external qubits, which
  may be interpreted as a probing apparatus for the inner Ising ring.
  Our results show that, in proximity of the quantum transitions of
  the Ising ring, either first-order or continuous, it is possible to
  put forward dynamic FSS frameworks which unveil peculiar scaling
  regimes, depending on the way in which the large-size limit is
  taken: either by fixing the number $n$ of probing qubits, or their
  interspace distance $b$.  In any case, the dependence of the various
  observables on $n$ can be reabsorbed into a redefinition of the
  quench parameter by a $\sqrt{n}$ prefactor.  We also address the
  role of a nearest-neighbor coupling between the external qubits.
\end{abstract}

\maketitle

% ========================= BODY =========================

\section{Introduction}
\label{intro}

The recent progress achieved by quantum simulators in carefully
controlling the dynamics of an increasing number of qubits has enticed
further theoretical studies on the coherent time evolution of quantum
correlations and energy exchanges of finite-size quantum systems (see,
e.g., Refs.~\cite{RV-21, Dziarmaga-10, PSSV-11, GAN-14, NC-book}).
On the one hand, a deeper understanding of the decoherence dynamics and
spreading of entanglement correlations is of fundamental importance,
both for quantum-information purposes and for the improvement of
energy conversion in complex networks~\cite{NC-book, LCCLCN-13}.  On
the other hand, the study of the energy storage among distinguished
parts of a quantum system, exploring different geometric schemes, is
relevant for quantum-thermodynamical purposes~\cite{GHRRS-16, VA-16},
as well as for the efficiency optimization of recently developed
quantum batteries~\cite{CPV-18}. In experiments with ultracold atoms,
such quantities can be also used to monitor the time evolution of a
finite amount of interacting qubits~\cite{SFSTT-20, BDZ-08}.

However, in any experimental setup, one is interested in bringing to
light the properties of the system under study, carefully minimizing
the perturbations ascribed to the devices employed to probe it.  As
the superposition principle drives disjoint parts of the same network
to be entangled during the time evolution, a collective framework in
which the system and the probing apparatus are treated on the same
footing is often necessary to correctly describe the evolution of the
overall setup~\cite{Zurek-03}.  In this regime, it is interesting to
find out how much information we may retrieve by considering the
coherent collective dynamics of the system and the probing apparatus
as a whole, but only looking at the quantum devices used to probe the
system under analysis.
Among the various schemes of composite systems that can be useful to
address this kind of issues, we mention the so-called {\em central
  spin} models, which have attracted a great deal of attention in the
recent years, both theoretically and experimentally (see,
e.g., Refs.~\cite{Zurek-82, HB-04, CPZ-05, QSLZS-06, RCGMF-07, CFP-07,
  BS-07, CP-08, Zurek-09, DQZ-11, NDD-12, SND-16, V-18, FCV-19, RV-19,
  ASSSCD-19, VCPPC-20, LSSWY-21, AM-21, NSC-21, RV-21}).

%%%%%%%%%%%%%%%%%%%%%%%%%%%%%%%%%%%%%%%%%%%%%%%%%%%%%%%%%%%%%%%%%%%%%%%%%%%%%%%%%%%%%%%%
\begin{figure}[!t]
  \centering
  \includegraphics[width=0.8\columnwidth, clip]{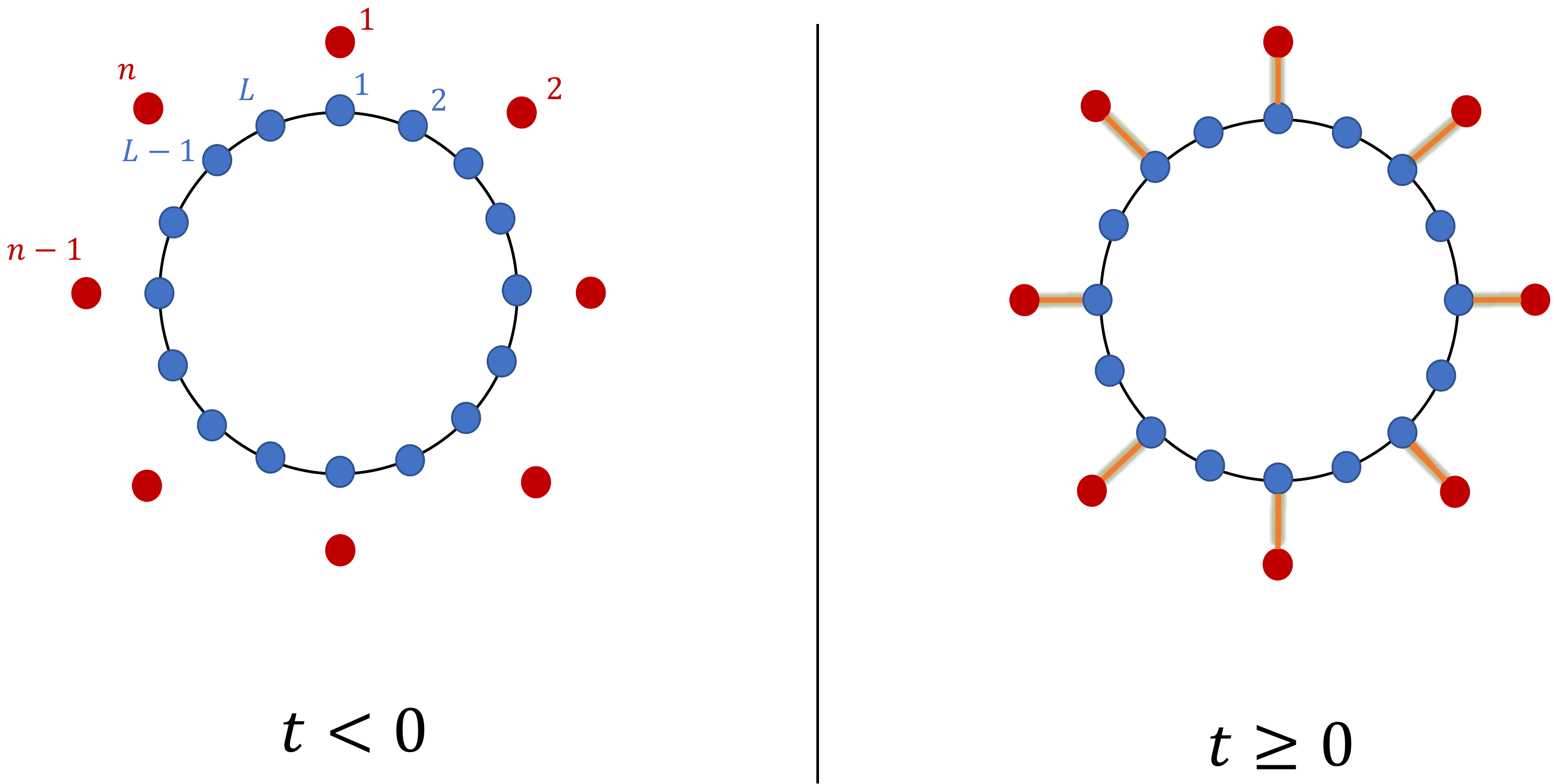}
  \caption{Skecth of the setup studied in this paper.  For times $t<0$
    (left side), the system is initialized and left in the
    ground-state configuration where ancillary spins ($\mathcal{A}$,
    red dots) belonging to the probing apparatus are initially
    decoupled from the Ising ring qubits ($\mathcal{I}$, blue dots).
    At time $t=0$, the ring and the external spins are suddenly
    coupled together (thick red lines) by quenching the Hamiltonian
    parameter $\kappa$ governing the $\mathcal{I}$-$\mathcal{A}$
    interaction, thus unitarily driving the global system out of
    equilibrium (right side).  In the figure, we set $L=16$, $n=8$,
    and $b=2$, see text.}
  \label{fig_quenchsunburst}
\end{figure}
%%%%%%%%%%%%%%%%%%%%%%%%%%%%%%%%%%%%%%%%%%%%%%%%%%%%%%%%%%%%%%%%%%%%%%%%%%%%%%%%%%%%%%%%

In this work we focus on an alternative lattice configuration, in the
class of the {\em sunburst} quantum models (for a sketch see
Fig.~\ref{fig_quenchsunburst}), in which many independent spins are
longitudinally coupled to a quantum Ising ring.  Specifically, we
consider a set of $L$ spin-1/2 quantum objects (qubits) disposed in a
ring geometry (blue dots), interacting through nearest-neighbor
Ising-like interactions.  We probe the system by means of $n$
equally-spaced qubits (being $b$ their interspace distance),
completely independent of each other, to which we will refer in the
following as the {\em ancillary} spins or simply the {\em external}
spins (red dots).  A thorough analysis of the equilibrium ground-state
properties of this apparatus has been recently performed in
Ref.~\cite{FRV-22}.  Here we are interested in investigating the
out-of-equilibrium dynamics triggered by a sudden quench of an
external parameter $\kappa$, which couples the system under study with
the monitoring devices.  When the monitoring devices act as a small
perturbation, leading to {\em soft} quenches, the out-of-equilibrium
quantum evolution is studied within dynamic finite-size scaling
frameworks, which turn out to be particularly effective to infer
dynamic scaling laws for finite-size systems~\cite{PRV-18, RV-21}.  In
particular, two different dynamic FSS limits can be defined at large
sizes, either by keeping the number $n$ of external qubits fixed or by
maintaining their interspace distance $b$ constant.  We have carried
out a careful FSS study of both limits, when the ring system is close
to either continuous or first-order quantum transitions (CQTs or
FOQTs, respectively) of the Ising ring. Moreover, we have considered
quenches in the Ising-disordered phase.  In any case, the dependence
of observables on $n$ can be always taken into account by a proper
rescaling of the parameter $\kappa$ by a $\sqrt{n}$ prefactor.  We
also discuss the many-body dynamics arising from {\em hard} quenches,
when the ancillary system introduces a relatively large amount of
energy with respect to the energy differences of the lowest levels of
the Ising ring. Finally, we address the role of a nearest-neighbor
coupling between the external qubits.

The paper is organized as follows. In Sec.~\ref{sec_sunburst_model} we
describe the setup of the model under analysis: the lattice
Hamiltonian, the quench protocol, and the observables addressed during
time evolution. In the next two sections we put forward a dynamic FSS
framework that holds for systems close to a CQT (Sec.~\ref{sec_CQT})
and to a FOQT (Sec.~\ref{sec_FOQT}), and support our theory with
numerical results based on exact diagonalization.  In
Sec.~\ref{sec_disordered_phase} we show some results at fixed $b$ in
the disordered phase.  In the last part of this work, we focus on hard
quenches, comparing our results with the ones
obtained when consecutive monitoring qubits also interact among
themselves (Sec.~\ref{sec_fixed_ring}).  Finally, our conclusions are
drawn in Sec.~\ref{sec_summary}.

\section{System setup}
\label{sec_sunburst_model}

\subsection{The lattice Hamiltonian}

We are going to address some properties of the unitary
out-of-equilibrium quantum dynamics of an Ising ring coupled to a set
of external qubits along the longitudinal axis, which is ruled by the
following global lattice Hamiltonian:
\begin{equation}
  \hat{H} = \hat{H}_{\mathcal I} + \hat{H}_{\mathcal A} + \hat{H}_{\mathcal IA}\,.
  \label{hamiltonian}
\end{equation}
The first term $\hat{H}_{\mathcal I}$ describes a standard
ferromagnetic Ising ring (${\mathcal I}$) in a transverse field, with
$L$ sites,
\begin{equation}
  \hat{H}_{\mathcal I} = -J\sum_{i=1}^L
  \hat{\sigma}^{(1)}_i \hat{\sigma}^{(1)}_{i+1}
  - g \sum_{i=1}^{L}\hat{\sigma}^{(3)}_i\,,
  \label{label_H_I}
\end{equation}
where $\hat{\sigma}^{(k)}_i$ (with $k=1,2,3$) denote the spin-1/2
Pauli matrices on the $i$th the lattice site, while
$\hat{\sigma}^{(k)}_{L+1}=\hat{\sigma}^{(k)}_1$, since we are
considering periodic boundary conditions (PBC).  In the following we
set $\hbar = 1$, $J=1$ as the energy scale, $a=1$ as the lattice
spacing, and take $g>0$ without loss of generality.

The Ising model is a paradigmatic quantum lattice system where
fundamental issues related to the physics of quantum many-body
networks can be thoroughly investigated, given our extended knowledge
of its quantum phases and correlations, both in the thermodynamic and
FSS limit~\cite{Sachdev-book, RV-21}.  The ground state exhibits a
zero-temperature CQT at $g=g_{\mathcal I}$ belonging to the
$(d+1)$-dimensional Ising universality class, separating a
ferromagnetic phase ($g<g_{\mathcal I}$) from a disordered phase
($g>g_{\mathcal I}$).  In the one-dimensional case, the quantum phase
transition is located at $g_{\mathcal I}=1$.  Close to the quantum
critical point (QCP), the system develops long-range correlations
$\xi$, diverging as $\xi\sim\abs{g - g_{\mathcal I}}^{-\nu}$, where
$\nu=y_g^{-1}=1$ and $y_g$ is the renormalization group (RG) exponent
associated with the difference $g-g_{\mathcal I}$.  The gap between
the ground-state energy and the first excited state
$\Delta_{\mathcal I}$ is instead suppressed as $\Delta_{\mathcal I}\sim\xi^{-z}$
close to the CQT (here $z$ denotes the dynamic critical exponent, $z=1$).
Consequently, for finite-size systems the gap is power-law suppressed
as $\Delta_{\mathcal I}(g=g_{\mathcal I}, L)=\pi/(2L)+O(L^{-2})$
at the QCP, whereas it vanishes exponentially at a FOQT when the
boundary conditions do not favor any of the phases
$\Delta_{\mathcal I}(g<g_{\mathcal I}, L)\approx e^{-c(g) \, L}$, as in the
case of PBC~\cite{CNPV-14, PRV-20}.

The second term $\hat{H}_{\mathcal A}$ in Eq.~(\ref{hamiltonian})
characterizes the energy of $n$ independent ancillary ($\mathcal{A}$)
spin-1/2 systems (qubits), acting as quantum monitoring devices for
the ring system under analysis,
\begin{equation}
  \hat{H}_{\mathcal A} = - \frac{\delta}{2}
  \sum_{i=1}^{n} \hat{\Sigma}^{(3)}_{i}\,,
  \label{hamiltonian_HA}
\end{equation}
where $\hat{\Sigma}^{(k)}_i$s  (with $k=1,2,3$)
still represent Pauli matrices on the
$i$th monitoring spin.  The coupling $\delta$ parameterizes the energy
variation defined by the flip of a single qubit, which is the
elementary excitation within the ensemble of the $n$ independent
qubits.

The last term $\hat{H}_{\mathcal IA}$ couples together the ring system
with the ancillary qubits along the longitudinal direction. In our
analysis, the number $n$ of equally-spaced spins is always
commensurate with the length $L$ of the Ising ring, so that each pair
of consecutive qubits is separated by a fixed distance $b=L/n$. It
reads as follows
\begin{equation}
  \hat{H}_{\mathcal IA} = - \kappa \sum_{i=1}^{n}
  \hat{\sigma}^{(1)}_{x=ib} \, \hat{\Sigma}^{(1)}_{i}\,,
  \label{hamiltonian_pairing}
\end{equation}
where $\kappa$ parameterizes the intensity of the coupling.

The model is characterized by a $\mathbb{Z}_2$ symmetry under
a global spin flip of both the ring and the ancillary spins,
identified by the parity operator
\begin{equation}
  \hat{P} = \bigg( \bigotimes_{i=1}^L\hat{\sigma}_i^{(3)}\bigg)
  \bigg( \bigotimes_{j=1}^n\hat{\Sigma}^{(3)}_j \bigg),
\end{equation}
which acts on the Pauli operators as
\begin{subequations}
  \begin{eqnarray}
  \hat{P} \, \hat{\sigma}^{(1)}_i \! & \to -\hat{\sigma}^{(1)}_i, \qquad \qquad
  \hat{P} \, \hat{\Sigma}^{(1)}_i \! & \to -\hat{\Sigma}^{(1)}_i \,,\\
  \hat{P} \, \hat{\sigma}^{(2)}_i \! & \to -\hat{\sigma}^{(2)}_i, \qquad \qquad
  \hat{P} \, \hat{\Sigma}^{(2)}_i \! & \to -\hat{\Sigma}^{(2)}_i \,,\\
  \hat{P} \, \hat{\sigma}^{(3)}_i \! & \to  \hat{\sigma}^{(3)}_i,
  \qquad \qquad \ \ \,
  \hat{P} \, \hat{\Sigma}^{(3)}_i \! & \to  \hat{\Sigma}^{(3)}_i\,.
\end{eqnarray}
\end{subequations}
We point out that the global Hamiltonian~(\ref{hamiltonian}) is left
invariant under a change in the sign of $\delta$ and/or $\kappa$,
after a proper redefinition of the Pauli matrices. Thus, without loss
of generality, in our analysis we will restrict ourselves to the cases
$\delta, \kappa \geq0$.

As mentioned in the introduction, the overall setup realizes the same
lattice configuration as the sunburst quantum Ising model, whose
equilibrium ground-state properties have been addressed in
Ref.~\cite{FRV-22}.  In a time-dependent perspective, the equilibrium
limit corresponds to an adiabatically slow dynamics for finite-size
systems.
In contrast, here we are going to focus on the out-of-equilibrium
decoherence properties and energy exchanges from the point of view of
the ancillary spins, aiming at perceiving the capability of these
probes to retrieve information on the Ising ring system under study,
and in particular its equilibrium quantum phases before the quench.
To this purpose, we consider two different dynamic FSS limits, depending
on the parameters we decide to maintain fixed with increasing the
lattice size $L$ of the ring~\cite{FRV-22}:
\begin{enumerate}
\item We keep the number $n=L/b$ of ancillary spins fixed while
  increasing $L$, thus the spins probing the ring get diluted in the
  FSS limit.  The bulk properties of the Ising ring system (such as
  the location of the QCP or the equilibrium phase diagram) are left
  unchanged by the presence of the probing devices.

\item We keep $b$ fixed, so that the number of external qubits
  increases linearly with the ring size $L$.  In this case, the qubits
  are expected to modify the bulk properties of the Ising ring (for
  example, the location of the ferromagnet-to-paramagnet QCP).
\end{enumerate}

\subsection{The quench protocol}
\label{sec_quench_protocol}

We now outline the sudden-quench protocol considered in the present
work (see Fig.~\ref{fig_quenchsunburst}). The Ising ring and the
ancillary spins are initially decoupled, i.e., with reference to
Eq.~(\ref{hamiltonian}), we start with $\kappa=0$ and choose the
initial state $\ket{\Psi_0}$ as the ground state of $\hat{H}_{\mathcal
  I}+\hat{H}_{\mathcal A}$:
\begin{equation}
  \ket{\Psi_0} = \ket{\psi_0} \otimes
  \bigg( \bigotimes_{i=1}^n\ket{\uparrow}_i \! \bigg)\,,
  \label{eq_groundstate_before_quench}
\end{equation}
where $\ket{\psi_0}$ is the ground state of the Ising ring and
$\otimes_{i=1}^n\ket{\uparrow}_i$ is the vector in the Hilbert space
$\mathcal{H}_{\mathcal A} = (\mathbb{C}^2)^n$ with all the $n$
auxiliary qubits pointing upward along the $(3)$-direction. At $t=0$
the system is suddenly driven out of equilibrium by quenching the
parameter $\kappa$ to a finite value: $0 \rightarrow \kappa$, such
that $\ket{\Psi_0}$ is no longer a Hamiltonian eigenstate.  Thus for
$t>0$ the global system evolves nontrivially in real time, according
to the usual Schr{\"o}dinger equation,
\begin{equation}
  \ket{\Psi(t)} = e^{-i\hat{H}t}\ket{\Psi_0}\,,
  \label{eq_schroedinger}
\end{equation}
where $\hat{H}$ is the total Hamiltonian reported in
Eq.~(\ref{hamiltonian}).

In the following discussion, we distinguish between two types of
sudden quench: a {\em soft} quench is related to a tiny change of the
parameter $\kappa$ (decreasing with $L$), so that the system stays
close to a quantum transition and thus excites only critical
low-energy modes. In contrast, a {\em hard} quench is not limited by
the above condition and typically involves the injection into the
system of an extensive amount of energy, in such a way that
also {\em high-energy} excitations are involved.

\subsection{Decoherence function and energy exchanges}

To characterize the scaling properties of the lattice model after the
quench, in the different phases and regimes considered, we focus on
the coherence and the energetic properties exhibited by the $n$
ancillary spins (subsystem $\mathcal{A}$).

First of all, we define the decoherence function $D_{\mathcal A}$
as~\cite{RV-21}
\begin{equation}
  D_{\mathcal A}(t) = 1 -
  \Tr_{\mathcal A}\big[ \hat{\rho}_{\mathcal A}^2(t)\big]\,,
  \label{def_decoherence}
\end{equation}
where $\hat{\rho}_{\mathcal A}\equiv\Tr_{\mathcal
  I}\big[\ket{\Psi}\bra{\Psi}\big]$ is the reduced density matrix of
the external qubits and $\Tr_{\mathcal{I}/\mathcal{A}}$ represents,
respectively, the trace over the Hilbert space of the ring system
($\mathcal{I}$) or that of the probing spins ($\mathcal{A}$).  This
observable quantifies the coherence properties of $\mathcal{A}$ and is
intimately related to the bipartite entanglement shared by the two
distinguished subsystems $\mathcal{I}$ (blue dots in
Fig.~\ref{fig_quenchsunburst}) and $\mathcal{A}$ (red dots in
Fig.~\ref{fig_quenchsunburst})~\footnote{More precisely, the
decoherence function $D$ is strictly connected with the second-order
R{\'e}nyi entropy, $S^{(2)}_{\mathcal A}(t) = - \log \, \Tr_{\mathcal
  A} \big[ \hat{\rho}_{\mathcal A}^2(t)\big]$, which is an
entanglement monotone.}.  Namely, it ranges from $D_{\mathcal A}=0$,
for a pure state, to $D_{\mathcal A} \simeq 1$, for a maximally
entangled state (in the latter case, deviations from the unit value
are due to finite Hilbert spaces $\mathcal{H}_{\mathcal A}$).  By
means of the Schmidt decomposition, it is easy to prove that, when the
global state $\ket{\Psi}$ is pure (equivalently, it can be expressed
in terms of a single wave function), the decoherence of the system is
equal to that of the $n$ external spins, i.e. $D_{\mathcal A} =
D_{\mathcal I}\equiv D$.  Here we deal with global pure states, being
our global system isolated, therefore from now on we will drop the
subscript ${}_{\mathcal A}$ (or ${}_{\mathcal I}$).

To address the energy-related and quantum thermodynamic properties of
the half-integer monitoring spins, we first define the work statistics
distribution $P(W)$ associated with the quantum quench presented in
Sec.~\ref{sec_quench_protocol}~\cite{TLH-07, CHT-11, RV-21}
\begin{equation}
  P(W) = \sum_{m, n} \delta \big[ W - (E_{f, n} - E_{i, m}) \big]
  \, \abs{\braket{n_f}{m_i}}^2 \, p_{i, m}\,,
\end{equation}
where $\{\ket{m_{i}},\ket{n_{f}}\}$ and $\{E_{i, m}, E_{f, n}\}$ are
the eigenstates and the eigenvalues of, respectively, the starting and
the final Hamiltonian (before and after the quench), and
$p_{i,m}=e^{-\beta E_{i,m}}/Z$ is the probability of finding the
initial state in the eigenstate $\ket{m_i}$ of the beginning
Hamiltonian, with $\beta\equiv1/T$ the inverse temperature and $Z={\rm
  Tr} \big[ e^{-\beta \hat H} \big]$ the partition function. At $T=0$
we have $p_{i,0}=1$ and $p_{i,\alpha}=0, \ \forall \alpha>0$.  The
momenta of the work distribution are defined by the integral
\begin{equation}
  \expval{W^m} \equiv  \int \dd{W} P(W) \, W^m\,.
\end{equation}
In particular, the first two momenta $\expval{W}, \expval{W^2}$ can be
easily addressed by evaluating a static expectation value on the
ground state (without the complete knowledge of the whole spectrum),
as they have an extremely simple expression in terms of
$\hat{H}_{\mathcal IA}$:
\begin{equation}
  \expval{W^m} = \mel{\Psi_0}{\hat{H}_{\mathcal IA}^m}{\Psi_0}\,,
  \quad \text{if} \ m=1, 2\,.
  \label{eq_momentum_W}
\end{equation}
The above relations do not hold for higher momenta of the
distribution, with $m>2$.  However, given the ground state
$\ket{\Psi_0}$ of Eq.~(\ref{eq_groundstate_before_quench}), one can
easily obtain two simple exact relations for the first two work
momenta done on the whole system at the quantum quench:
\begin{equation}
  \expval{W}=0\,,\qquad \expval{W^2}=n\kappa^2\,.
  \label{workav}
\end{equation}

After quenching at $t=0$, the energy of the global sunburst system is
conserved along the evolution for $t>0$. However, we have significant
exchange of energy between the two subsystems. We define the
energy-exchange distribution (i.e. the quantum heat) of the qubits
associated with two different energy measurements of the subsystem
$\mathcal{A}$, the first one at $t=0$ and the next one at a generic
time $t>0$~\cite{CHT-11, RV-21},
\begin{equation}
  P_{\mathcal A}(U, t) \equiv \sum_{n, \alpha}
  \delta\big[U - (E_{\mathcal{A}_\alpha}-E_{\mathcal{A}_0})\big] \,
  \abs{\!\mel{n, \alpha}{e^{-i\hat{H}t}}{\Psi_0}}^2\,,
  \label{def_heat_distribution}
\end{equation}
where $n$ and $\alpha$ represent, respectively, the energy levels of
the ring system and the ones associated with the external spins before
the quench.  In particular, a simple relation holds for the first
momentum $\expval{U}$ of the distribution, i.e.
\begin{equation}
  \expval{U}(t) = \mel{\Psi(t)}{\hat{H}_{\mathcal A}}{\Psi(t)}
  - \mel{\Psi_0}{\hat{H}_{\mathcal A}}{\Psi_0}\,.
  \label{def_prob_distribution_U}
\end{equation}
We remark that the above observable quantifies the average energy
stored into the set of $n$ half-integer spins, as soon as the
$\mathcal{I}$ and $\mathcal{A}$ subsystems are decoupled at a given
time $t$.  With this respect, the quantum quench protocol we address
can be seen, in a different perspective, as the refilling protocol of
$n$ discharged quantum batteries connected to a charger Ising ring
system.

\section{Quantum quenches at the CQT}
\label{sec_CQT}

\subsection{Finite number of qubits in the FSS regime}
\label{CQT_finite_n_FSSregime}

In this section, we present the dynamic FSS framework derived for soft
quenches: at $t = 0$, the system is suddenly driven out of equilibrium
by switching on the coupling $\kappa$, such that the global system
remains close to a QCP ($g\approx g_{\mathcal I}$) and the pairing
Hamiltonian $\hat{H}_{\mathcal IA}$ acts as a small perturbation of
$\hat{H}_{\mathcal I}+\hat{H}_{\mathcal A}$.  We begin our analysis by
keeping the number $n$ of external qubits fixed in the large-$L$ limit
[FSS limit (i)].

As a starting point, we recap some recent FSS results
put forward for the equilibrium ground-state properties for the
sunburst quantum Ising model~\cite{FRV-22}, in which $\kappa$ is
different from zero and does not depend on time, and then introduce
our scaling ansatz required to characterize the various dynamic
regimes, basing our arguments on FSS and RG arguments.
The natural scaling variable associated with the transverse field
strength $g$ is defined analogously as in the standard quantum Ising
model close to the CQT, therefore it reads:
\begin{subequations}
  \label{def_WAK}
\begin{equation}
  G = (g - g_{\mathcal I}) \, L^{y_g}, \qquad y_g=1\,,
  \label{def_W}
\end{equation}
where $y_g=1/\nu$ is its RG dimension ($\nu=1$ in one dimension).  On
the basis of energy scales considerations, a common ansatz to provide
us with another working scaling variable is given by the ratio of the
two relevant low-energy scales $\delta$, which characterizes the
energy gap of the ancillary qubits close to the critical point, and
$\Delta_{\mathcal I}(g_{\mathcal I}, L)$, which represents the
finite-size gap of the Ising ring at the QCP.  Therefore we consider
the scaling variable
\begin{equation}
  A = \frac{\delta}{\Delta_{\mathcal I}(g_{\mathcal I},L)}
  \sim \delta \, L^z, \qquad z=1\,,
  \label{def_A}
\end{equation}
where $z$ is the dynamic critical exponent.  The scaling variable
associated with the quench parameter $\kappa$ is expected to play a
role similar to that of a quench generated by Ising symmetry-breaking
defects $\sum_{i=1}^n\hat{D}_i=-\kappa
\sum_{i=1}^n\hat{\sigma}_{x=ib}^{(1)}$ at a CQT. For a finite number
$n$ of external spins, the RG dimension of the Hamiltonian parameter
$\kappa$ at the Ising fixed point is given by
$y_\kappa=(2+z-d-\eta)/2=7/8$~\cite{FRV-22a}. Therefore, we define the
corresponding scaling variable as~\cite{FRV-22}
\begin{equation}
  K = \kappa \, L^{y_\kappa}, \qquad y_\kappa=7/8\,.
  \label{def_K}
\end{equation}
\end{subequations}

Then, the FSS limit is obtained by taking the large-size limit ($L \to
\infty$), while keeping constant the values of the above mentioned
scaling variables $G, A$, and $K$ in Eqs.~(\ref{def_WAK}), therefore
requiring $\delta \sim L^{-z}$ and $\kappa \sim L^{-y_\kappa}$.  For
instance, a natural ansatz for the scaling behavior of the equilibrium
ground-state decoherence function $D= 1- {\rm Tr}_{\mathcal A} \, [
  \hat \rho^2_{\mathcal A}]$ in the FSS limit and close to the CQT of
the Ising ring is given by the scaling equation~\cite{FRV-22}
\begin{equation}
  D(n, g, \delta, \kappa, L) \approx \mathcal{D}(n, G, A, K)\,,
  \label{FSS_static_decoherence}
\end{equation}
with $D(n,g,\delta,0,L)=\mathcal{D}(n, G, A, 0) = 0$.  Analogous
scaling ansatzes can be put forward for other quantities and
observables, such as the gap of the global system, the two-point
correlation function of the order parameter, as well as its
susceptibility, the correlation length, etc \ldots~\cite{FRV-22}.  These
have been carefully verified by means of extensive numerical
simulations for finite-size systems in which all the Hamiltonian
parameters $g$, $\delta$, and $\kappa$ have been properly rescaled
with $L$ according to the above definitions of scaling
variables~\cite{FRV-22}.

We now generalize the above picture to a dynamic scenario, induced by
a sudden switch on of the coupling between the ancillary and Ising
subsystems, from zero to a nonzero value $\kappa$. We consider a soft
quench, in such a way that the coupling parameter $\kappa$ after the
quench decreases with $L$, keeping the scaling variables $K$ fixed.
In general, dynamic behaviors exhibiting nontrivial time dependences
are expected to require another scaling variable associated with the
time variable $t$, defined as~\cite{RV-21}
\begin{equation}
  \Theta = t \, \Delta_{\mathcal I}(g_{\mathcal I},L) \sim t \, L^{-z},
  \qquad z=1\,.
  \label{def_Theta_scaling_variable}
\end{equation}
The FSS limit for $L \to +\infty$ is obtained by keeping constant $G,
A, K$ in Eqs.~(\ref{def_WAK}), and also the time variable $\Theta$ in
Eq.~(\ref{def_Theta_scaling_variable}).  Therefore, in this
circumstance, we conjecture that observables should generally satisfy
universal relations, depending on their scaling dimensions and
arguments.
For instance, a natural ansatz for the scaling behavior of the
decoherence function $D$ of Eq.~(\ref{def_decoherence}), in the FSS
limit, is given by the scaling equation
\begin{equation}
  D(n, g, \delta, \kappa, t, L) \approx \mathcal{D}(n, G, A, K,
  \Theta)\,,
  \label{FSS_decoherence}
\end{equation}
which generalizes Eq.~(\ref{FSS_static_decoherence}) to the dynamic
case, by simply adding a further dependence on the time scaling
variable $\Theta$.  We conjecture an analogous scaling ansatz for the
average energy $\expval{U}$ stored into the auxiliary qubits
[cf.~Eq.~(\ref{def_prob_distribution_U})]:
\begin{equation}
  \expval{U}(n, g, \delta, \kappa, t, L)
  \approx L^{-z} \mathcal{E}(n, G, A, K,\Theta)\,,
  \label{FSS_energy}
\end{equation}
assuming that further analytical contributions are
suppressed~\cite{RV-21}.

As we shall see, the approach to the asymptotic dynamic FSS when
keeping $n$ finite is generally characterized by leading $O(L^{-7/4})$
corrections.  They arise from the analytic expansion of the non-linear
scaling field associated with the quench parameter $\kappa$, i.e.,
$u_\kappa\approx \kappa + c\kappa^3+...$ (where the quadratic term
vanishes, due to the symmetric properties of the Ising ring parameter
$\kappa$), whose third-order term gives rise to corrections decaying
as $O(L^{-7/4})$~\cite{FRV-22}. We also mention that also subleading
$O(L^{-2})$ corrections are generally present in one-dimensional
systems with PBC, while systems with boundaries are generally subject
to $O(L^{-1})$ corrections~\cite{CPV-14}.

%%%%%%%%%%%%%%%%%%%%%%%%%%%%%%%%%%%%%%%%%%%%%%%%%%%%%%%%%%%%%%%%%%%%%%%%%%%%%%%%%%%%%%%%
\begin{figure}[!t]
  \centering
  \includegraphics[width=0.95\columnwidth, clip]{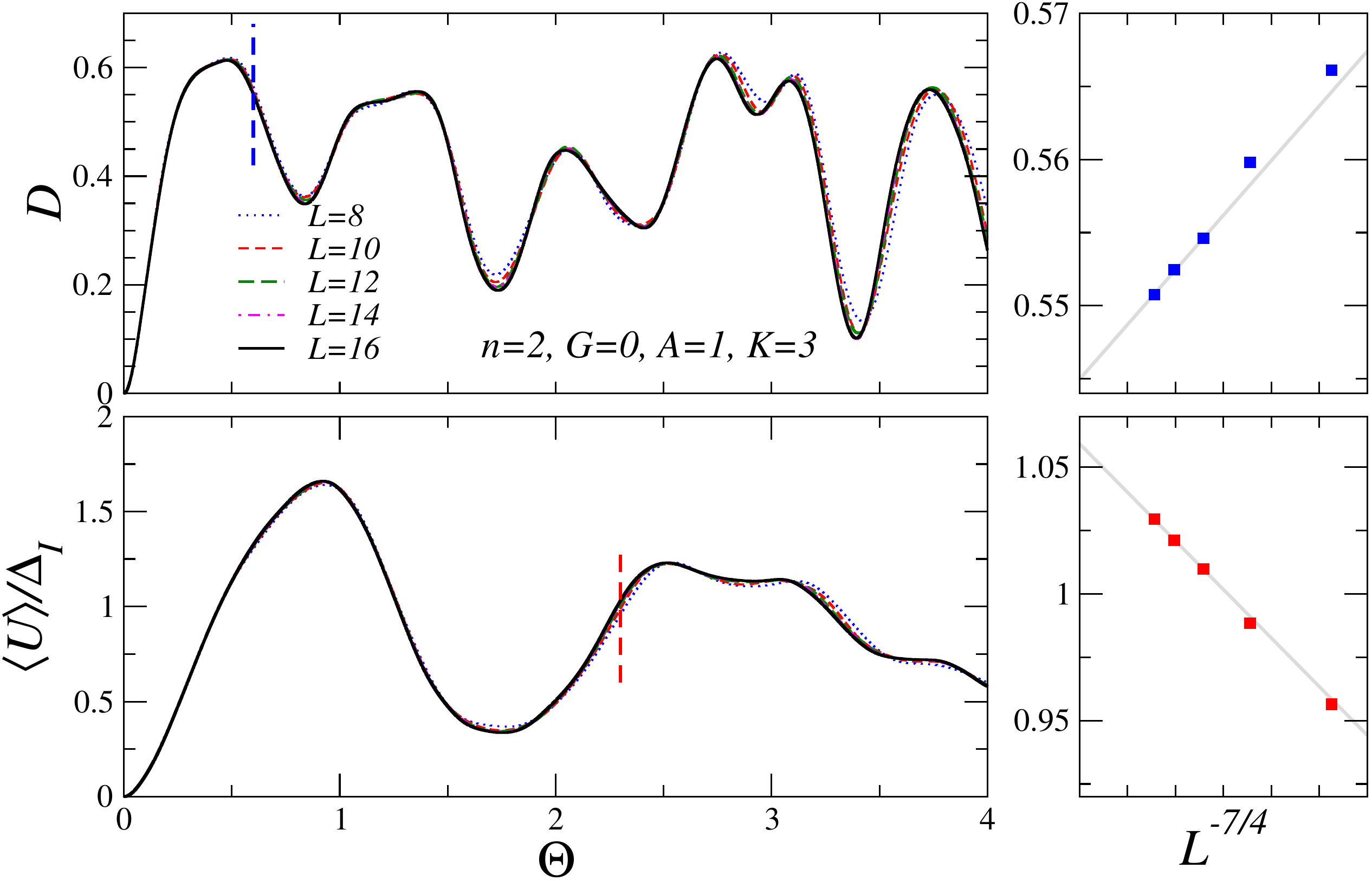}
  \caption{Analysis of the decoherence function $D$ (top panels) and
    of the rescaled energy stored in the external qubits $\expval{U}
    /\Delta_{\mathcal I}$ (bottom panels) in the FSS limit keeping the
    number $n$ of external spins fixed.  Left panels: the quantities
    versus the rescaled time variable $\Theta$.  The different curves
    are obtained for various Ising-ring lengths up to $L=16$ (and
    fixed $n=2$, $G=0$, $A=1$, $K=3$), starting from the initial state
    $\ket{\Psi_0}$ of Eq.~(\ref{eq_groundstate_before_quench}).  Right
    panels: the corresponding scaling corrections for fixed
    $\Theta=0.6$ and $2.3$, respectively) are shown to be consistent
    with a decay $\sim L^{-7/4}$ (straight lines are drawn to guide to
    the eye).}
  \label{fig_CQT}
\end{figure}
%%%%%%%%%%%%%%%%%%%%%%%%%%%%%%%%%%%%%%%%%%%%%%%%%%%%%%%%%%%%%%%%%%%%%%%%%%%%%%%%%%%%%%%%

To verify the above dynamic FSS relations~(\ref{FSS_decoherence})
and~(\ref{FSS_energy}) at the CQT, in Fig.~\ref{fig_CQT} we present
some numerical results obtained by means of exact diagonalization
and $4^{\text{th}}$-order
Runge-Kutta techniques to find the coherent time evolution by
integrating the Schr{\"o}dinger equation in
Eq.~(\ref{eq_schroedinger})~\footnote{We always used a discrete time
step of the order of $\delta t= (1 \divisionsymbol 3) \times 10^{-3}$
We have checked that the interval $\delta t$ is sufficiently small to
allow us to neglect systematic errors arising from the discretization
of the temporal evolution.}.  As visible from the nice data collapse
of the various curves obtained for different ring lengths $L$, the
presented results are in agreement with our dynamic FSS theory.

\subsection{FSS in the large-$n$ limit}
\label{sec_largen}

\begin{figure}[!t]
  \centering
  \includegraphics[width=0.95\columnwidth, clip]{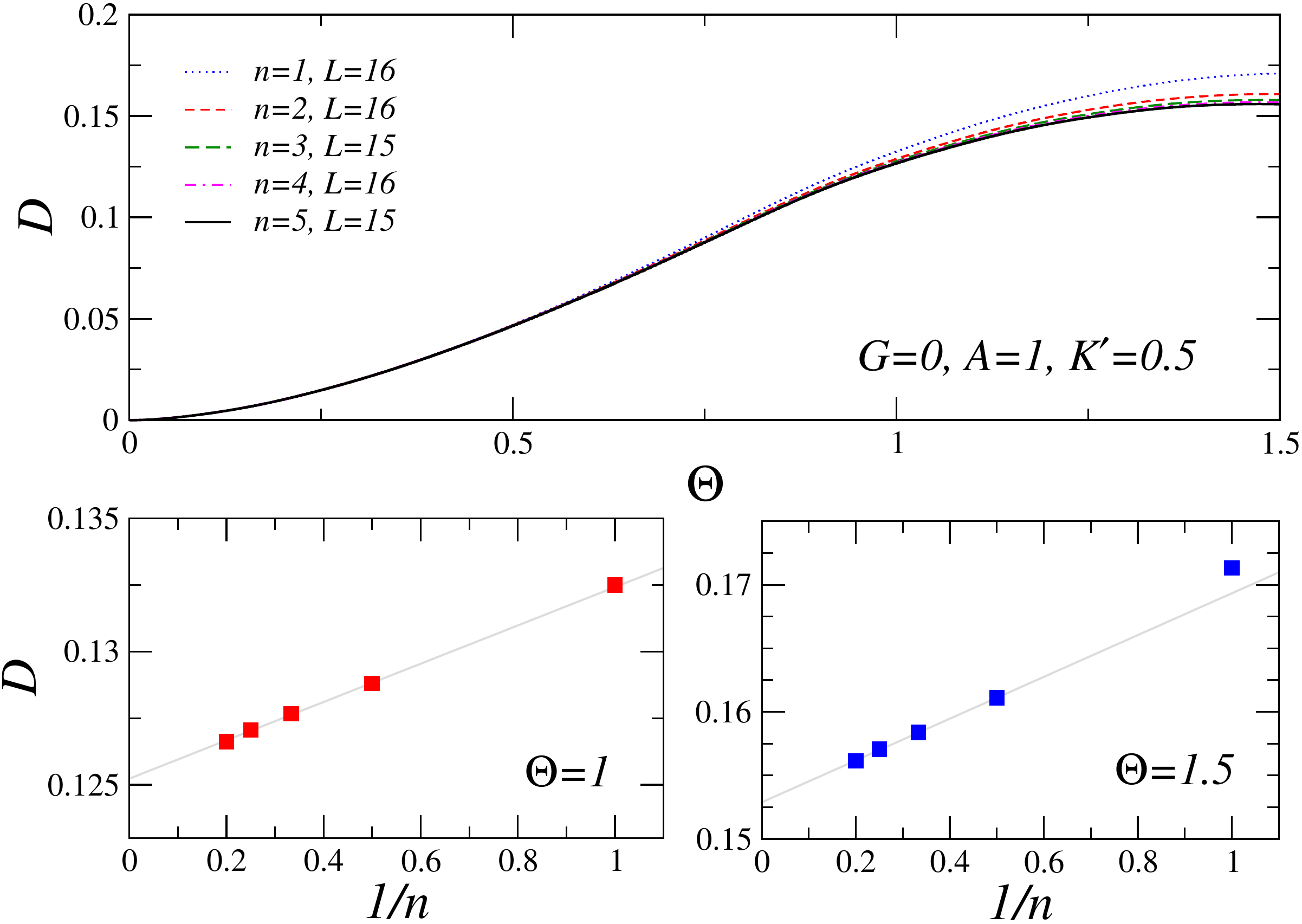}
  \caption{Top panel: the decoherence function $D$ versus $\Theta$,
    for several values of $n$ at the largest lattice sizes at our
    disposal, for fixed $G=0$, $A=1$, $K^\prime=0.5$.  We observe data
    collapse with increasing $n$, within scaling corrections that get
    asymptotically suppressed as $n^{-1}$, as shown for $\Theta=1$
    (bottom left panel) and $\Theta = 1.5$ (bottom right panel). We
    emphasize that each point in the bottom panels is obtained, for
    fixed $n$, by means of large-size extrapolations ($L \to \infty$,
    at fixed $n$), assuming $\sim L^{-7/4}$ scaling
    corrections. Straight lines are drawn to guide to the eye.}
  \label{fig_sqrtn_CQT}
\end{figure}

It is also interesting to note that, for sufficiently large $n$, the
dependence on the number of ancillary spins $n$ can be absorbed into a
redefinition of the quench parameter $K$, for sufficiently large $n$:
\begin{equation}
  K^\prime = \sqrt{n} \, K\,.
  \label{def_sqrtn_K}
\end{equation}
Note the large-$n$ limit considered here should be performed after the
$L\to\infty$ limit.  In particular, the quantum model under
investigation exhibits appreciable hints of a $\sqrt{n}$-law already
for relatively small values of $n \sim 3 \divisionsymbol 5$, as
reported in Fig.~\ref{fig_sqrtn_CQT} for the decoherence function $D$.
The asymptotic regime appears to be rapidly approached, with $O(1/n)$
scaling corrections.  The square-root dependence on $n$ of the quench
parameter $K^\prime$ turns out to be the result of the collective
(independent) behavior of the ancillary spins, which are indirectly
coupled only through the Ising ring. Other observables we have
obtained by means of exact diagonalization techniques, as the ratio
$\expval{U}/\Delta_{\mathcal I}(g_{\mathcal I}, L)$, obey the same
large-$n$ FSS relation introduced for $K^\prime$ (not shown).

%%%%%%%%%%%%%%%%%%%%%%%%%%%%%%%%%%%%%%%%%%%%%%%%%%%%%%%%%%%%%%%%%%%%%%%%%%%%%%%%%%%%%%%%
%%%%%%%%%%%%%%%%%%%%%%%%%%%%%%%%%%%%%%%%%%%%%%%%%%%%%%%%%%%%%%%%%%%%%%%%%%%%%%%%%%%%%%%%

\subsection{Qubits at fixed distance $b$ in the FSS regime}
\label{sec_bfixed}

We carry on our discussion by considering the case of soft quenches in
which the external spins are placed at fixed distance $b$ in proximity
of a CQT, thus in the large-$n$ limit [FSS limit (ii)].  As working
hypothesis, we follow the reasoning for the equilibrium
scenario~\cite{FRV-22} and suppose the theory put forward in the
previous section for large $n$ [in particular, the $\sqrt{n}$-law of
  Eq.~(\ref{def_sqrtn_K})] to survive also in this different FSS
limit.  The most natural guess for the scaling variable associated
with the quench strength $\kappa$ is given in terms of the new scaling
variable $\widetilde{K}$, obtained by replacing $\sqrt{n}\to \sqrt{L/b}$
in the definition of $K'$, cf. Eq.~(\ref{def_sqrtn_K}), i.e.
\begin{equation}
  \widetilde{K} \equiv \sqrt{n} \, K = \kappa \, \frac{L^{11/8}}{b^{1/2}}\,.
  \label{def_widetilde_K}
\end{equation}

According to this hypothesis, the previous dynamic FSS relations,
cf. Eqs.~(\ref{FSS_decoherence}) and~(\ref{FSS_energy}), should be
modified by removing the $n$ dependence in the corresponding scaling
functions and substituting $K$ with $\widetilde{K}$.  For example, for
the decoherence function $D$, we conjecture that
\begin{equation}
  D(b, g, \delta, \kappa, t, L)
  \approx \mathcal{D}(A, \widetilde{K},\Theta)\,.
  \label{FSS_decoherence_2}
\end{equation}
Analogous relations are expected to hold also for other
observables, as for the energy stored into the monitoring spins
$\expval{U}$.
Note that the dependence of $D$ on $n$ (which is an extensive energy
by definition) is completely encapsulated in the attenuation of the
quench intensity $K$ (being $\widetilde{K}$ fixed in the FSS regime)
by a factor $\sqrt{n}\sim \sqrt{L}$.

\begin{figure}[!t]
  \centering
  \includegraphics[width=0.95\columnwidth, clip]{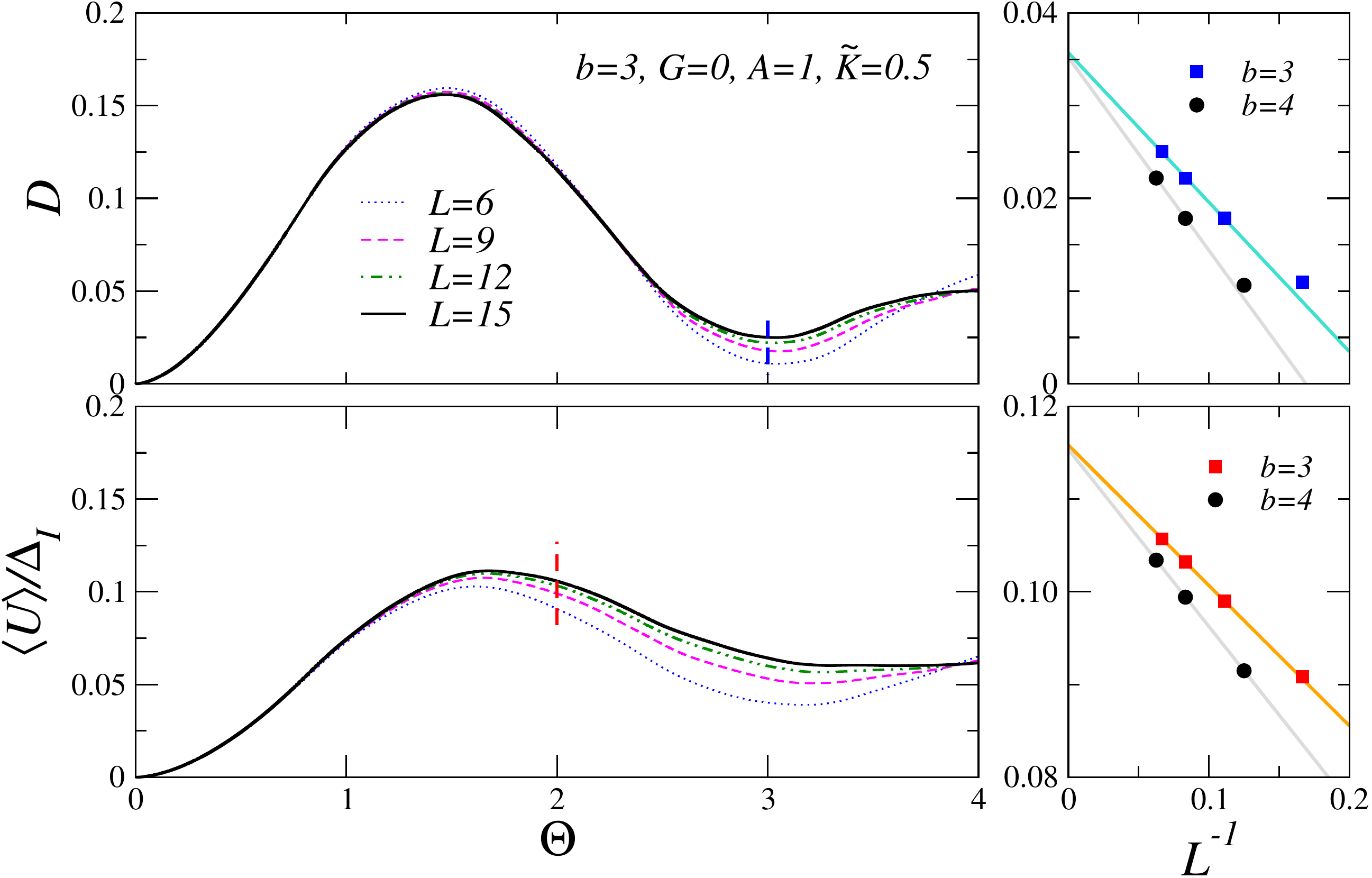}
  \caption{Scaling of the decoherence function $D$ and the
    RG-invariant ratio $\expval{U}/\Delta_{\mathcal I}$ versus
    $\Theta$ (top and bottom left panels, respectively), for fixed
    $b=3$ and $G=0$, $A=1$, $\widetilde{K} = 0.5$.  Right panels: the
    corresponding data for $b=3$ and $b=4$ (not shown on the left), at
    fixed $\Theta=3$ and $2$ respectively, are consistent with each
    other, within scaling corrections suppressed as $\sim
    L^{-1}$. Straight lines are drawn to guide the eye.}
  \label{fig_CQT_b_scaling}
\end{figure}

The above conjecture has been verified numerically in
Fig.~\ref{fig_CQT_b_scaling}, for a fixed distance $b=3$ and for $G=0,
A=1$, and $\widetilde{K}=0.5$.  In this figure the decoherence
function $D$, together with the ratio $\expval{U}/\Delta_{\mathcal
  I}$, show a nice scaling in terms of the variable $\Theta$, thus
confirming our hypothesis.  We mention that results for $b=3$ and $4$
are consistent with each other, within scaling corrections that decay
roughly as $\sim L^{-1}$. Such scaling corrections may be related to
the $O(1/n)$ corrections observed in the large-$n$ limit considered in
Sec.~\ref{sec_largen}, since $n\sim L$ at fixed $b$. To leave the
figure easily readable, in the left panels we report the time behavior
only for the $b=3$ data, however it is worth mentioning that the
scaling of the $b=4$ data with increasing lattice sizes is very
similar to that shown for $b=3$, at all times $\Theta<4$.

%%%%%%%%%%%%%%%%%%%%%%%%%%%%%%%%%%%%%%%%%%%%%%%%%%%%%%%%%%%%%%%%%%%%%%%%%%%%%%%%%%%%%%%%
%%%%%%%%%%%%%%%%%%%%%%%%%%%%%%%%%%%%%%%%%%%%%%%%%%%%%%%%%%%%%%%%%%%%%%%%%%%%%%%%%%%%%%%%

\subsection{Qubits at fixed distance $b$ and fixed $\delta$}
\label{sec_CQT_finite_delta}

We now consider the dynamics of the quantum Ising ring model, with
many independent spins attached, in the hard-quench limit: at $t = 0$,
the system is suddenly driven out of equilibrium by switching on the
coupling $\kappa$ from zero to a fixed finite value, in such a way
that an extensive amount of energy is injected.  We fix the spacing
distance parameter $b$ and the excitation energy $\delta$, examining
different ring lengths $L$. Note also that when $\delta>0$ the energy
scale of the ancillary system is much larger than the gap of the
critical Ising ring. In this case, the dynamic FSS frameworks
presented in the previous section do not hold anymore, since the hard
quench drives the system out of the critical domain. Therefore,
the resulting dynamics should be little sensitive to the
value $g$ of the parameter controlling the Ising criticality.

\begin{figure}[!t]
    \centering
    \includegraphics[width=0.95\columnwidth, clip]{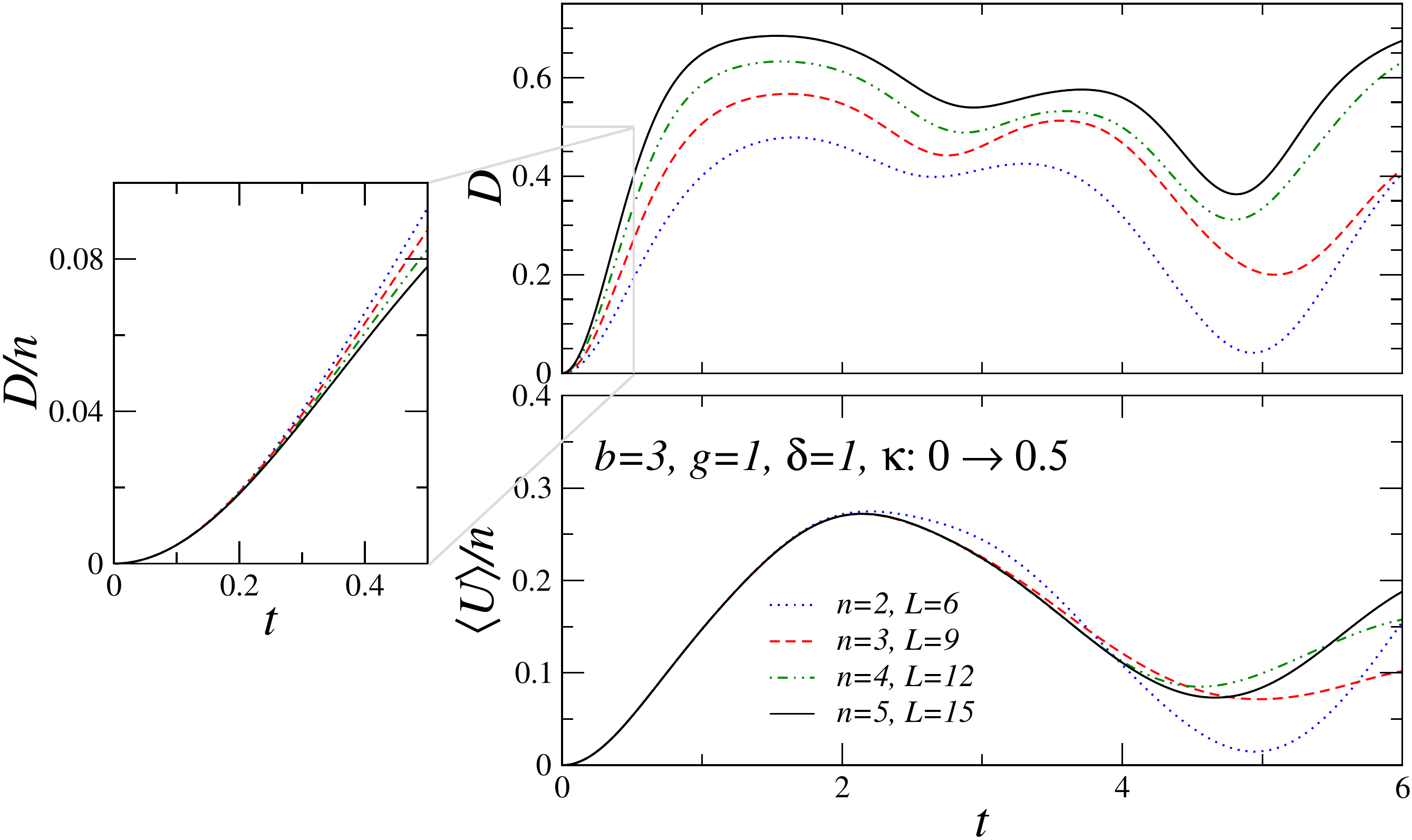}
    \caption{Top panel: the decoherence function $D$ versus
      $t$. Different curves are for various values of $n$ and $L$,
      while fixing $b=3$, setting $g=\delta=1$, and quenching $\kappa$
      from $0$ to $0.5$. We note that $D$ generally increases with $n$
      and, for $t\ll1$, its growth is proportional to the number $n$
      of qubits attached to the ring system, as shown in the inset on
      the left panel that zooms in on the domain $t\in[0, 0.5]$ (note
      that on the y-axis we show $D/n$). Bottom panel: the energy stored per
      unit qubit $\expval{U}/n$ versus the time $t$.  The first peak
      at $t\approx2$ is stationary with increasing $n$, consistently
      with an extensive scaling of the energy.}
    \label{fig_manyn_fixed_delta}
\end{figure}

In Fig.~\ref{fig_manyn_fixed_delta} we show results for fixed
$g=\delta=1$ and for a quench intensity $\kappa:0\to 0.5$. Differently
from the FSS regime analyzed in the previous section, the energy
stored per unit qubit is $\sim O(1)$, thus scaling extensively with
the number of ancillary qubits $n$ (see the peak in the top panel).
For small times $t\lesssim0.2$, we also observe a linear growth with
$n$ in the decoherence function $D$, for lattice sizes up to $L=15$.

%%%%%%%%%%%%%%%%%%%%%%%%%%%%%%%%%%%%%%%%%%%%%%%%%%%%%%%%%%%%%%%%%%%%%%%%%%%%%%%%%%%%%%%%
%%%%%%%%%%%%%%%%%%%%%%%%%%%%%%%%%%%%%%%%%%%%%%%%%%%%%%%%%%%%%%%%%%%%%%%%%%%%%%%%%%%%%%%%

\section{Quantum quenches along the FOQT line}
\label{sec_FOQT}

\subsection{Finite number of qubits in the FSS regime}
\label{FOQT_FSS}

We now analyze the model in proximity of the FOQT line ($g<g_{\mathcal
  I}$). For the sake of simplicity, we start from the case with a
fixed number $n$ of ancillary spins [FSS limit (i)].  We stress again
that, in the FSS and thermodynamic limit, a finite number of external
spins cannot change the bulk properties of the ring system, and in
particular its equilibrium quantum phases.  To begin with, we present
the scaling variables employed across this section to derive dynamic
FSS relations for quantum quenches at FOQTs.

Still relying our FSS arguments on the basis of the relevant energy
scales associated with each subsystem, we can first introduce the
analogue of the scaling variables~(\ref{def_A}) and~(\ref{def_K}) for
systems close to FOQTs, in the static case.  Namely, we define the
variable~\cite{FRV-22, RV-21}
\begin{subequations}
\begin{equation}
  Y_\delta = \frac{\delta}{\Delta_{\mathcal I}(g, L)}\,,
  \label{def_Y_delta}
\end{equation}
as the ratio between the energy of the elementary excitation of a
single qubit and the gap of the finite-size Ising ring
$\Delta_{\mathcal I}(g, L)$ at size $L$ and for a transverse field
$g<g_{\mathcal I}$.  Similarly to Ising models at the FOQT driven by a
longitudinal magnetic field, we may also introduce the scaling
variable $Y_\kappa$ associated with the variable
$\kappa$,~\cite{FRV-22, CNPV-14}
\begin{equation}
  Y_\kappa = \frac{\kappa}{\Delta_{\mathcal I}(g, L)}\,,
  \label{def_Y_kappa}
\end{equation}
\end{subequations}
which determines the {\em softness} of the quantum quench.  To extend
the FSS framework to the dynamic scenario, a further scaling variable
associated to the time $t$ is required, analogously as was done in
Eq.~(\ref{def_Theta_scaling_variable}).  Namely,~\cite{PRV-18}
\begin{equation}
  \Theta = t \; \Delta_{\mathcal I}(g, L)\,,
  \label{def_Gamma}
\end{equation}
where $\Delta_{\mathcal I}(g,L)$ now denotes the gap of the Ising-ring
system for $g<g_{\mathcal I}$.  Therefore, differently from CQTs in
the FSS limit, data collapse of observables at FOQTs is expected to be
recovered at time scales that substantially depend on the way in which
the gap $\Delta_{\mathcal I}$ closes with $L$: generally, these may
range from power-law to exponentially large times, according to the
different boundary conditions~\cite{RV-21}.  In our specific case
with a PBC geometry, we expect an exponential scaling
$\Delta_{\mathcal I}(g<g_{\mathcal I},L) \propto g^L$.

Relations analogous to those presented at the CQTs are still formally
valid in proximity of FOQTs, after properly replacing the
corresponding scaling variables.  For example, the dynamic FSS
relations related to the decoherence function $D$
[cf.~Eq.~(\ref{FSS_decoherence}), for systems close to the CQT] is now
expected to scale as
\begin{subequations}
\label{FSS_FOQT}
\begin{equation}
  D(n, g, \delta, \kappa, t, L) \approx
  \mathcal{D}(n, g, Y_\delta, Y_\kappa,\Theta)\,,
\end{equation}
while the energy stored into the auxiliary qubits
[cf.~Eq.~(\ref{FSS_energy}) at the CQT]
should read 
\begin{equation}
  \expval{U}(n, g, \delta, \kappa, t, L)
  \approx \Delta_{\mathcal I}(g, L) \;\,
  \mathcal{E}(n, g, Y_\delta, Y_\kappa, \Theta)\,.
\end{equation}
\end{subequations}
The analogous counterparts for static conditions with a constant and
finite $\kappa$ (i.e., without time dependence induced by changes of
the parameter $\kappa$) have been numerically verified in
Ref.~\cite{FRV-22}.  Here we rather focus on numerically testing
our above dynamic FSS framework for soft quenches close to the FOQT
line $g<g_{\mathcal I}$.

%%%%%%%%%%%%%%%%%%%%%%%%%%%%%%%%%%%%%%%%%%%%%%%%%%%%%%%%%%%%%%%%%%%%%%%%%%%%%%%%%%%%%%%%
\begin{figure}[!t]
  \centering
  \includegraphics[width=0.95\columnwidth, clip]{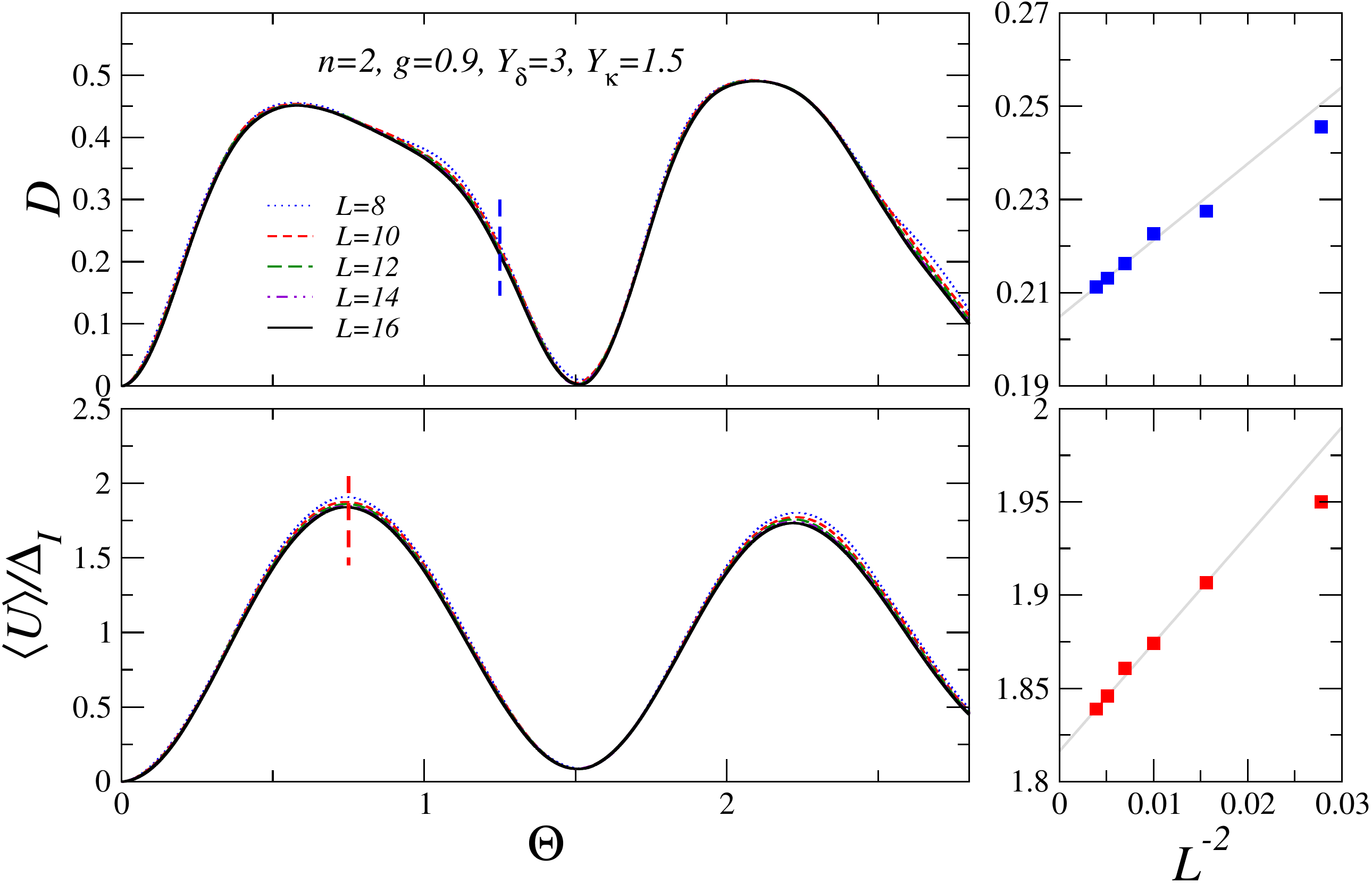}
  \caption{Top left panel: the decoherence function $D$ in terms of
    the rescaled time $\Theta$ at the FOQT for $g=0.9$, with
    $Y_\delta=3, Y_\kappa=1.5$, and for fixed $n=2$.  Bottom left
    panel: the ratio $\expval{U}/\Delta_{\mathcal I}$ versus $\Theta$.
    In the corresponding right panels we report the same data as a
    function of $L$, for fixed $\Theta = 1.25$ and $0.75$, showing
    that scaling corrections are substantially consistent with a decay
    $\sim L^{-2}$.  Straight lines are drawn to guide the eye.}
  \label{fig_FOQT_nfixed}
\end{figure}
%%%%%%%%%%%%%%%%%%%%%%%%%%%%%%%%%%%%%%%%%%%%%%%%%%%%%%%%%%%%%%%%%%%%%%%%%%%%%%%%%%%%%%%%

Specifically, the dynamic FSS laws in Eqs.~(\ref{FSS_FOQT}) for
systems close to FOQTs are supported by the data in
Fig.~\ref{fig_FOQT_nfixed}, which shows an excellent data collapse for
the various curves at different ring sizes $L$, and for rescaled times
smaller than $\Theta<2.8$ (left panels).  Scaling corrections appear
to be suppressed roughly as $\sim L^{-2}$, with increasing $L$ (right
panels).  We have checked that the above FSS predictions are verified
numerically also for a different value of $g=0.8$, by considering the
case for fixed $n=2$ and ring sizes up to $L=12$ (not shown).

\begin{figure}[!t]
  \centering
  \includegraphics[width=0.95\columnwidth, clip]{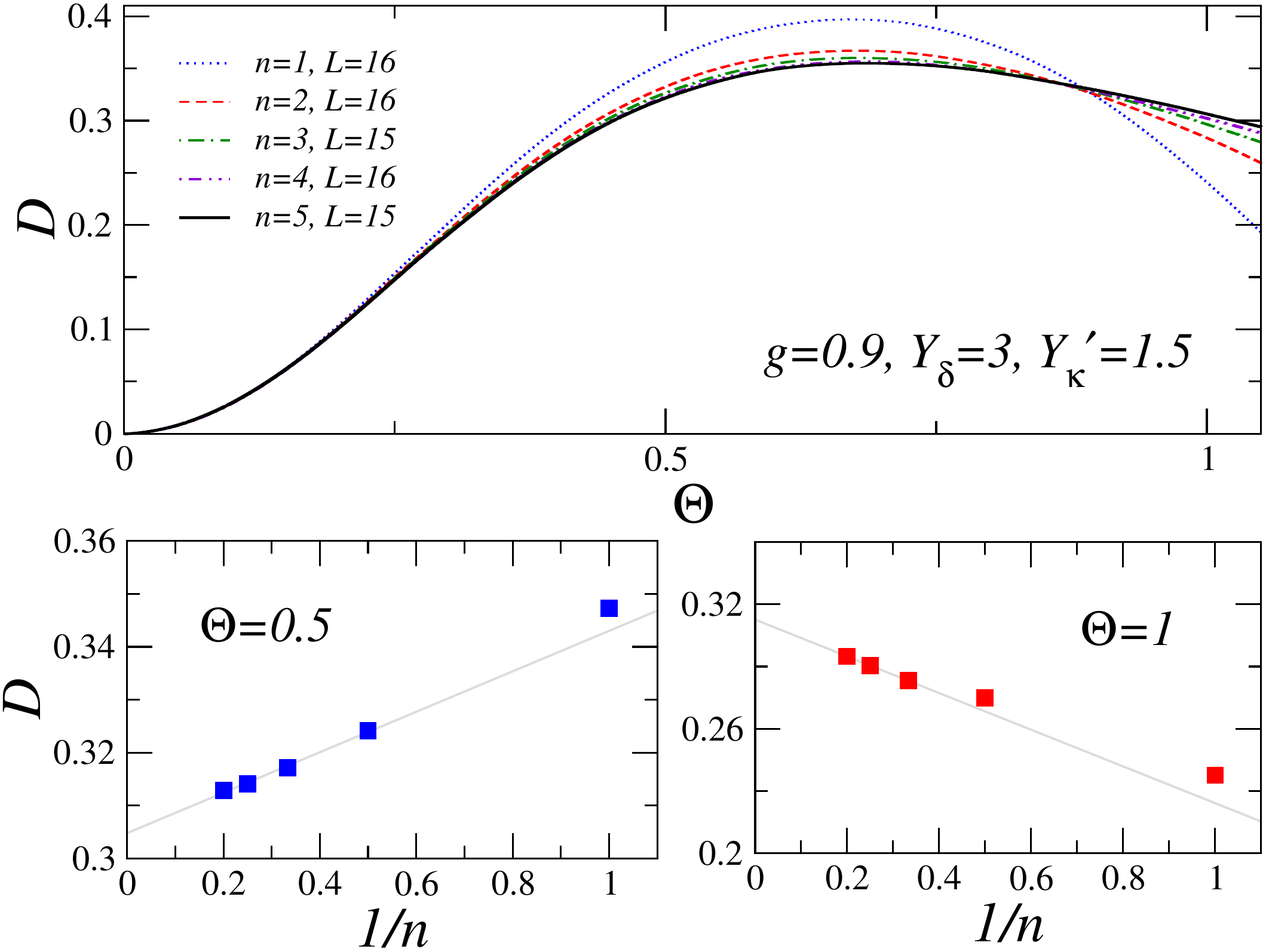}
  \caption{Top panel: the decoherence function $D$ versus $\Theta$,
    for several values of $n=1, \ldots ,5$ for the largest lattice
    sizes available with our exact diagonalization methods.  Scaling
    corrections at $\Theta=0.5$ (bottom left panel) and $1$ (bottom
    right panel) are consistent with a decay $\sim L^{-1}$.  Straight
    lines are drawn to guide the eye.  Here we stay close to the FOQT
    at $g=0.9$, fixing $Y_\delta=3$ and $Y_\kappa^\prime=1.5$.}
  \label{fig_manyn_FOQT}
\end{figure}

Also at FOQTs we observed that the dependence of observables on the
number of ancillary spins could be absorbed into a redefinition of the
variable $Y_\kappa$ as
\begin{equation}
  Y^\prime_\kappa = \sqrt{n} \, Y_\kappa\,.
  \label{def_Y_kappa_prime}
\end{equation}
To this extent, in Fig.~\ref{fig_manyn_FOQT} we present result for the
decoherence function $D$ for the largest lattice sizes at our
disposal, for several values of $n$.  For fixed $\Theta = 0.5$ (bottom
left) and $\Theta = 1$ (bottom right panel) we observe a power-law
convergence in the number of external spins as $\sim
n^{-1}$~\footnote{For each $n$, we have performed large-size
extrapolations assuming $L^{-2}$ scaling corrections, by fitting an
ansatz function $D(L)=a + b/L^{2}$}.

%%%%%%%%%%%%%%%%%%%%%%%%%%%%%%%%%%%%%%%%%%%%%%%%%%%%%%%%%%%%%%%%%%%%%%%%%%%%%%%%%%%%%%%%
%%%%%%%%%%%%%%%%%%%%%%%%%%%%%%%%%%%%%%%%%%%%%%%%%%%%%%%%%%%%%%%%%%%%%%%%%%%%%%%%%%%%%%%%
\subsection{Qubits at fixed distance $b$ in the FSS regime}
\label{sec_FOQT_bfixed}

\begin{figure}[!t]
  \centering
  \includegraphics[width=0.95\columnwidth, clip]{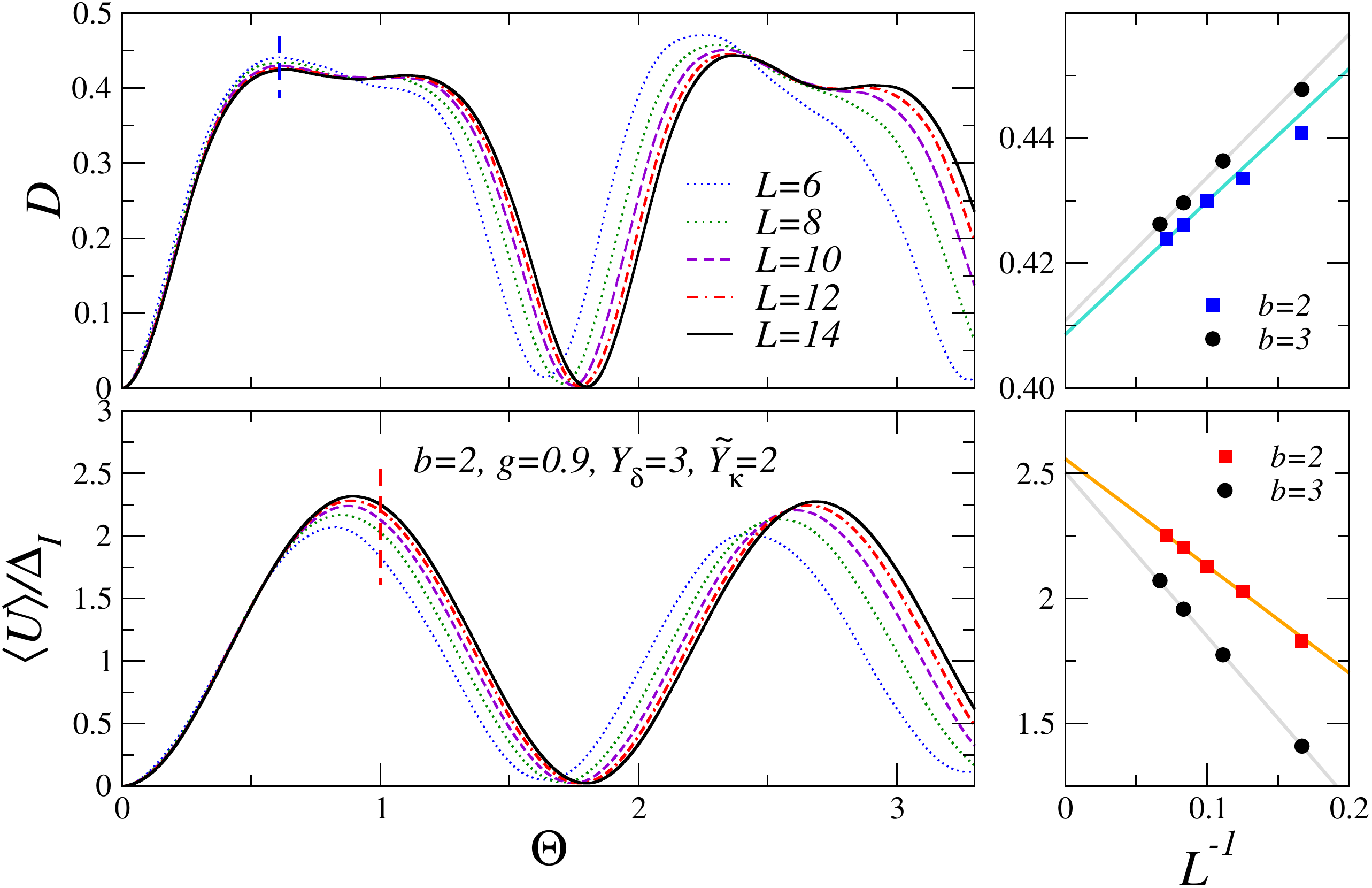}
  \caption{Top panels: the decoherence function $D$ versus the
    rescaled time $\Theta$ (left) and the inverse size (right). Bottom
    panels: same as in the top panels, but for the stored energy
    $\expval{U}$ rescaled with the gap $\Delta_{\mathcal I}(g=0.9, L)$
    of the finite-size Ising ring.  The scaling corrections shown in
    the right panels, evaluated for the top and bottom panels
    respectively at $\Theta=0.6$ and $1$, are in good agreement with a
    decay $L^{-1}$ (straight lines are drawn to guide the eye).  We
    fix $g=0.9$, $b=2$, and $Y_\delta=3$, $\widetilde{Y}_\kappa=2$.
    However, similarly to Fig.~\ref{fig_CQT_b_scaling}, in the insets
    we also show some data associated with a different value of $b=3$,
    to emphasize the independence of the observables on $b$ in the
    large $L$ limit, while keeping $\widetilde{Y}_\kappa=\kappa
    L^{1/2}/[b^{1/2}\Delta_{\mathcal I}(g, L)]$ fixed.}
    \label{fig_scaling_FOQT_2b}
\end{figure}

Following the results carried out in the previous section, we now
extend our dynamic FSS framework at FOQTs to the domain when $b$ is
fixed, thus the number of quantum probes increases with the ring size
$L$ [FSS limit (ii)]. In this limit, analogously to what was done in
Eq.~(\ref{def_widetilde_K}) close to a CQT, we recognize as a relevant
scaling variable for the quench intensity the ratio~\cite{FRV-22}
\begin{equation}
  \widetilde{Y}_K = \sqrt{n} \,
  Y_K = \frac{\kappa L^{1/2}}{\Delta_{\mathcal I}(g, L) \, b^{1/2}}\,.
  \label{def_K_tilde}
\end{equation}
With reference to the previous section, dynamic FSS relations can be
obtained by replacing $Y_K$ with $\widetilde{Y}_K$, e.g., in
Eqs.~(\ref{FSS_FOQT}).  Our scaling hypotheses are checked numerically
in Fig.~\ref{fig_scaling_FOQT_2b} for lattice sizes up to $L=14$,
within scaling corrections that appear to decay as $\sim L^{-1}$,
which should be again related to the $O(1/n)$ corrections found in the
large-$n$ limit at fixed $n$.

%%%%%%%%%%%%%%%%%%%%%%%%%%%%%%%%%%%%%%%%%%%%%%%%%%%%%%%%%%%%%%%%%%%%%%%%%%%%%%%%%%%%%%%%
%%%%%%%%%%%%%%%%%%%%%%%%%%%%%%%%%%%%%%%%%%%%%%%%%%%%%%%%%%%%%%%%%%%%%%%%%%%%%%%%%%%%%%%%

We skip the discussion of the hard-quench limit starting from the
ferromagnetic phase with $g<g_{\mathcal I}$, since we do not find any
relevant difference to that found when performing the quench from the
QCP.  In the zero-temperature equilibrium phase diagram of the
sunburst quantum Ising model, it was indeed conjectured that, when $b$
is fixed, the critical line characterizing the breaking of the
$\mathbb{Z}_2$ global symmetry shifts to higher values
$g_c(\kappa)>g_{\mathcal I}$ for any $\kappa \neq 0$~\cite{FRV-22}.
Since the hard-quench protocol we consider always
pushes the system away from criticality into the ordered phase, we do
not find any remarkable difference between quenches performed when
$g<g_{\mathcal I}$ or $g=g_{\mathcal I}$.

\section{Quantum quenches from the disordered phase}
\label{sec_disordered_phase}

Let us now present some numerical results for the dynamics of our
lattice system, when it is driven out of equilibrium from the disordered
phase ($g>g_{\mathcal I}$), fixing the distance $b$ between two
consecutive external spins.  Guided by the $\sqrt{n}$-behavior put
forward in the dynamic FSS at CQTs and FOQTs, and by the equilibrium
properties of the sunburst quantum Ising model~\cite{FRV-22}, we first
address the dynamics of the system in the disordered phase with
rescaled quench intensity $\kappa: 0\to 1/\sqrt{n}$.

%%%%%%%%%%%%%%%%%%%%%%%%%%%%%%%%%%%%%%%%%%%%%%%%%%%%%%%%%%%%%%%%%%%%%%%%%%%%%%%%%%%%%%%%
\begin{figure}[!t]
  \centering
  \includegraphics[width=0.95\columnwidth, clip]{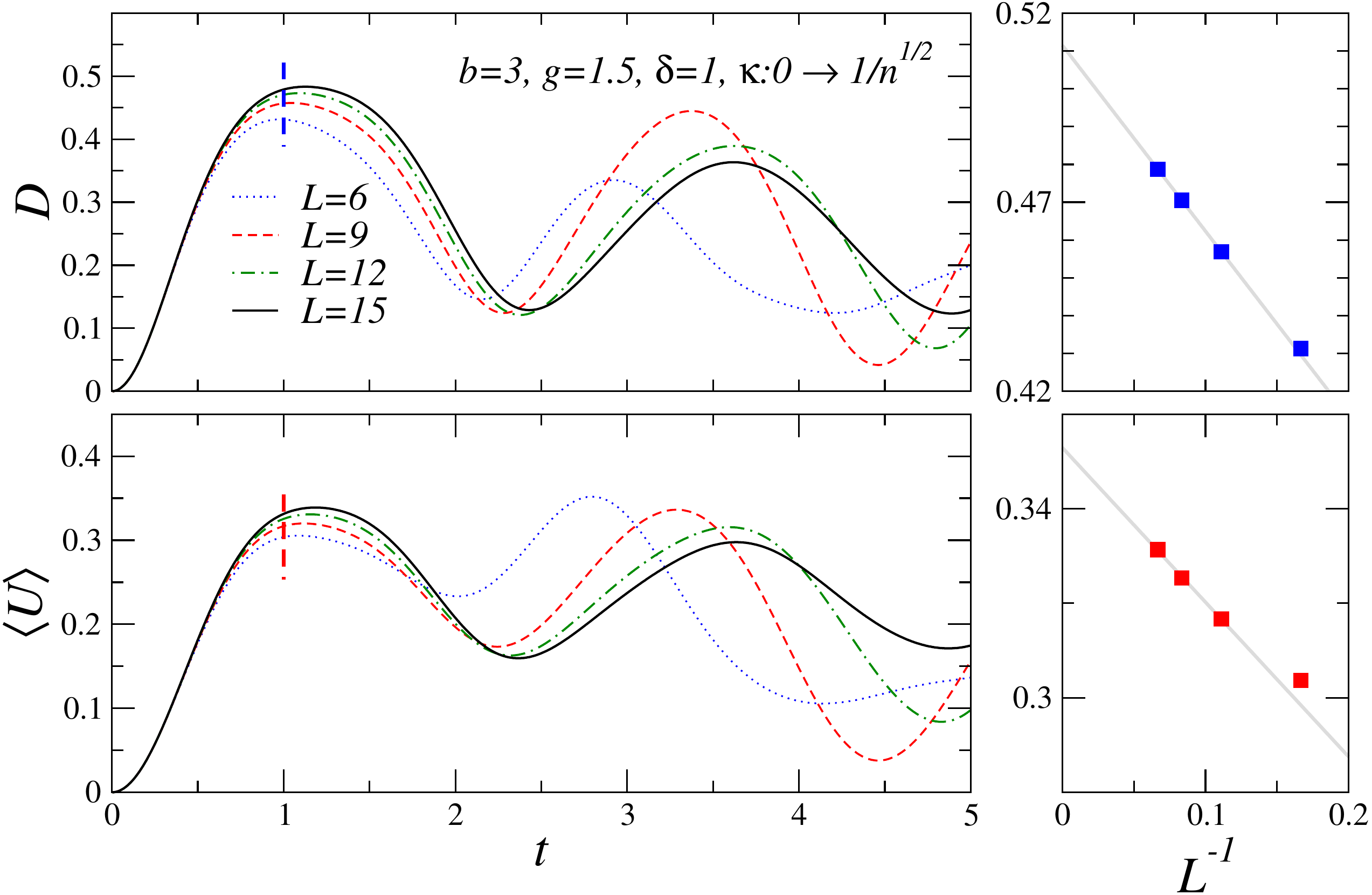}
  \caption{Results for quantum quenches starting from the disordered
    phase.  We keep $b=3, g=1.5, \delta=1$ fixed, and rescale the
    intensity of the quench with $n$, such that
    $\kappa:0\to1/\sqrt{n}$. Top left panel: the decoherence function
    $D$ versus $t$.  Bottom left panel: the energy $\expval{U}\propto
    n$ stored into the ancillary qubits versus $t$.  Scaling
    corrections, shown in the corresponding right panels for $t=1$,
    are consistent with a decay $\sim L^{-1}$ (straight lines are
    drawn to guide the eye).}
  \label{fig_disordersqrtn}
\end{figure}
%%%%%%%%%%%%%%%%%%%%%%%%%%%%%%%%%%%%%%%%%%%%%%%%%%%%%%%%%%%%%%%%%%%%%%%%%%%%%%%%%%%%%%%%

\begin{figure}
    \centering
    \includegraphics[width=0.95\columnwidth, clip]{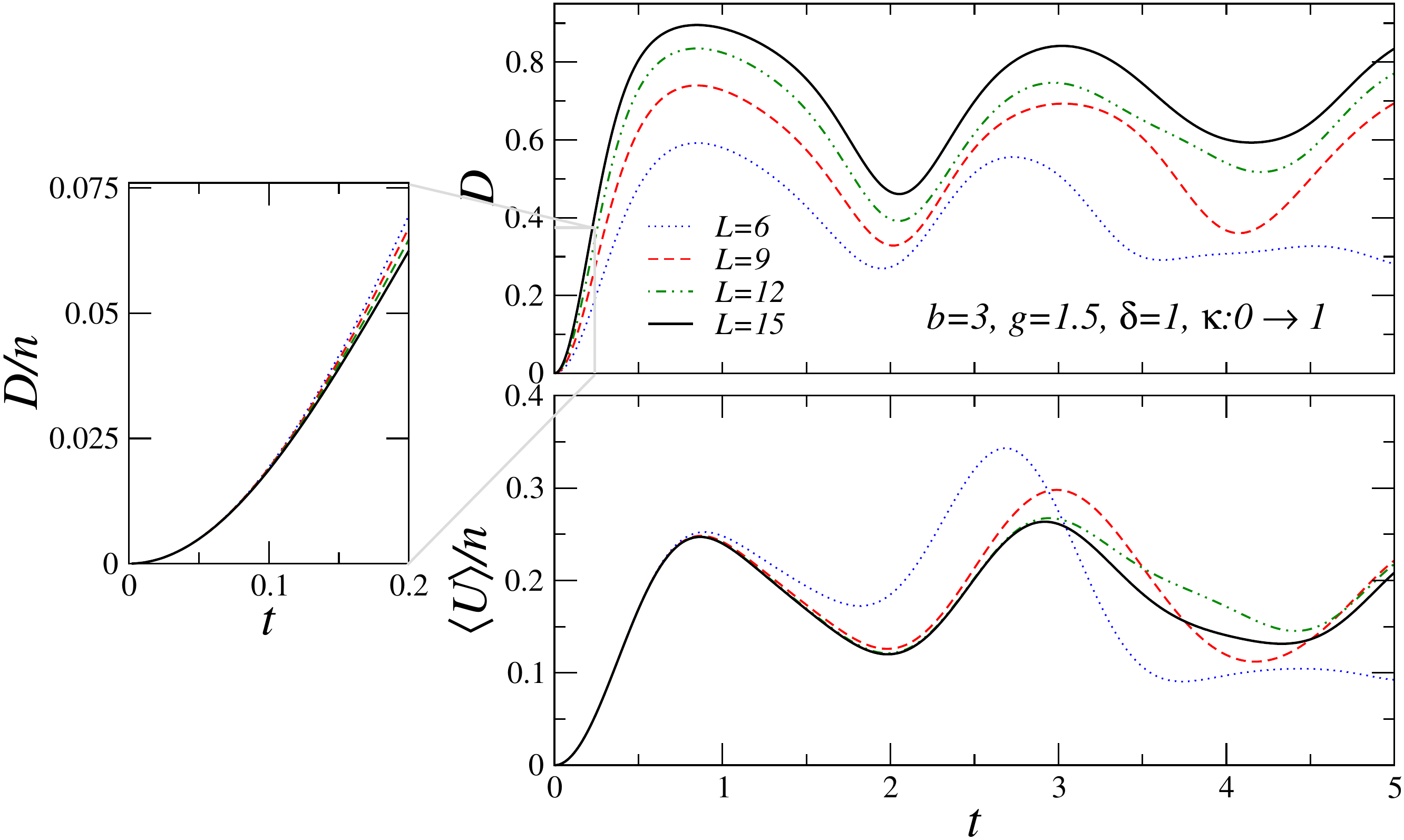}
    \caption{Results for quantum quenches starting from the disordered
      phase ($g=1.5$). We maintain fixed $b=3$, $\delta=1$ and perform
      a quench for $\kappa:0\to1$. The decoherence function generally increases
      with increasing the number of ancillary spins $n$. The inset shows that
      such function increases linearly with $n$ at small times $t\lesssim0.2$.
      Bottom panel: The maximum energy per unit spin encapsulated
      in the external qubits is roughly independent of $n$.}
    \label{fig_disordered_kfixed}
\end{figure}

In Fig.~\ref{fig_disordersqrtn} we present results for fixed $g=1.5$
and $\delta=1$.  We remember the reader that the typical fluctuations
associated with the global work done on the overall system at the
quench scale as $\sqrt{\expval{W^2}}=\kappa \sqrt{n}$, so that it
remains constant in this limit.  Therefore, in
this regime, the coupling between the ring and the external spins
decreases as the lattice size increases. Our results suggest that, for
small times $t\lesssim 1$, the peak of both the decoherence function
$D$ and the energy stored into the probes $\expval{U}$ approach a
constant behavior for large $L$ (note that the latter quantity
increases linearly with the lattice size $\expval{U}\propto n \langle
\hat \Sigma^{(3)} \rangle$).
We also present some results for fixed quench intensity $\kappa$ and $g=1.5$
in Fig.~\ref{fig_disordered_kfixed}. As expected, in this second case,
the decoherence function $D$ generally increases with the number $n$
of external spins (at small times it depends on $n$ as $D\propto n$,
as shown in the correspondent inset) and the first peak of $\expval{U}$
scales extensively with the number of ancillaries.

%%%%%%%%%%%%%%%%%%%%%%%%%%%%%%%%%%%%%%%%%%%%%%%%%%%%%%%%%%%%%%%%%%%%%%%%%%%%%%%%%%%%%%%%
%%%%%%%%%%%%%%%%%%%%%%%%%%%%%%%%%%%%%%%%%%%%%%%%%%%%%%%%%%%%%%%%%%%%%%%%%%%%%%%%%%%%%%%%

\section{Quenches with Ising ring of fixed length}
\label{sec_fixed_ring}

Finally, we analyze some examples of hard quenches when the length $L$
of the Ising ring is fixed, but the number of external probes vary,
which is perhaps the most relevant case for quantum-metrology purposes
or for setting a hypothetical quantum-battery device.  To this extent,
the focus is on the role of $n$ in determining the dynamics of
observables (in particular the energy $\expval{U}$) when the ancillary
qubits which probe the ring are independent from each other. To remark
this fact, we compare our results with the analogous ones obtained by
coupling consecutive nearest-neighbor external qubits along the
transverse direction.  As an exploratory study, we consider the
following extension of the ancillary Hamiltonian~(\ref{hamiltonian_HA})
\begin{equation}
  \hat{H}^\prime_{\mathcal A} = \hat{H}_{\mathcal A} - J_\Sigma
  \sum_{i=1}^{n} \hat{\Sigma}^{(3)}_{i} \hat{\Sigma}^{(3)}_{i+1},
  \label{eq_H_p_kinetic_ancillary}
\end{equation}
where PBC are intended in the second term, thus preserving the
translational invariance of the lattice model, modulo $b$.  In the
following we present some results for the case $J_\Sigma=1$.

Note that the pre-quench state $\ket{\Psi_0}$ in
Eq.~(\ref{eq_groundstate_before_quench}) is still the ground state of
the global system with $\kappa=0$, independently of the presence of
the new interaction term of Eq.~(\ref{eq_H_p_kinetic_ancillary}).  Of
course, the post-quench unitary dynamics will change, and thus also
the average energy stored into the qubits:
\begin{equation}
  \expval{U} = \mel{\Psi(t)}{\hat{H}^\prime_{\mathcal A}}{\Psi(t)}
  - \mel{\Psi_0}{\hat{H}^\prime_{\mathcal A}}{\Psi_0}\,.
  \label{eq_heat_kinetic}
\end{equation}

%%%%%%%%%%%%%%%%%%%%%%%%%%%%%%%%%%%%%%%%%%%%%%%%%%%%%%%%%%%%%%%%%%%%%%%%%%%%%%%%%%%%%%%%
\begin{figure}[!t]
  \centering
  \includegraphics[width=0.95\columnwidth, clip]{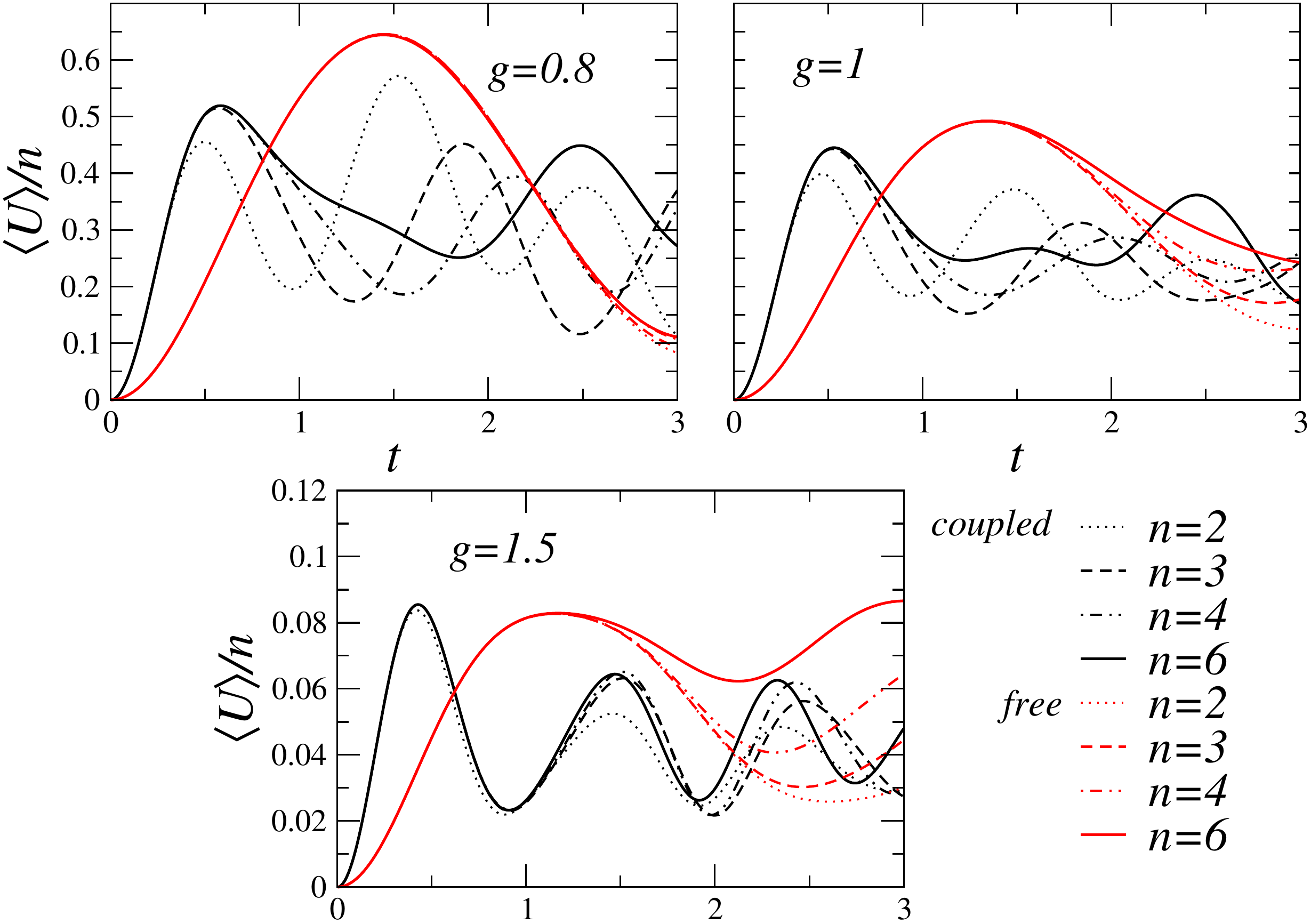}
  \caption{The average energy per unit spin stored into an external
    qubit versus the time $t$, for both cases of independent external
    spins ({\em free} in the legend) and of external spins which
    interact according to Eq.~(\ref{eq_H_p_kinetic_ancillary}) ({\em
      coupled} in the legend). Here we consider $L=12$ and
    vary $n$ as indicated in the legend, fix $\delta=1$, and perform a
    quench of $\kappa$ from $0$ to $1$ for several values of $g$.
    There is no striking difference between the two quenches shown at
    $g=0.8$ and $1$: the {\em free} and the {\em coupled} model can
    absorb roughly the same amount of energy
    $\expval{U}/n\approx0.5$. The capability of energy absorption
    worsen in both models for $g=1.5$ (note that the $y$ scale in the
    last graph is different from the upper panels).  }
  \label{fig_L_fixed}
\end{figure}
%%%%%%%%%%%%%%%%%%%%%%%%%%%%%%%%%%%%%%%%%%%%%%%%%%%%%%%%%%%%%%%%%%%%%%%%%%%%%%%%%%%%%%%%

In Fig.~\ref{fig_L_fixed} we present some results for fixed
$\delta=1$, quench intensity $\kappa:0\to 1$ and ring length $L=12$
(which offers several combinations of $b$, commensurate to the ring
length) to be investigated. Looking at the peaks in the energy stored
for all the three values of the coupling $g$ we have considered, each
one representing a different phase of the Ising ring, we first note an
extensive scaling of the energy with the number $n$ of qubits: In
particular, the maximum energy stored into a single spin is roughly
the same for $g=0.8$ and $1$, $\expval{U}/n \approx 0.45
\divisionsymbol 0.50$, whereas it appears strongly suppressed to
$\expval{U}/n \approx 0.08$ for both models, when $g=1.5$).  However,
it is also worth noting that the charging of the qubits that are {\em
  coupled} is generally significantly faster than that obtained when
the external qubits are independent from each other.  This last fact
holds for all values of $g$ shown in Fig.~\ref{fig_L_fixed}, as well
as the extensiveness with $n$ of the energy stored $\expval{U}$. We
have not found evidence of striking differences between the energy
stored into the probing apparatus of the two models. However, we
should say that we have not done an extended study of the optimal
parameters in this respect.

\section{Summary and conclusions}
\label{sec_summary}

We have studied the post-quench unitary dynamics of an ensemble of
quantum spin-1/2 objects arranged in a sunburst geometry: namely, at a
given reference time, some of the $L$ spins of a transverse-field
quantum Ising ring are suddenly and locally coupled to a set of $n$
independent external qubits ($n \leq L$), along the longitudinal
direction, in such a way to maintain a residual translation invariance
and the Ising $\mathbb{Z}_2$ symmetry.  The system exhibits various
quantum phases separated by first order (FOQTs) or continuous (CQTs)
quantum transitions, which are controlled by the various Hamiltonian
parameters as the Ising transverse field $g$, the energy scale
$\delta$ of the external qubits, and the interaction strength $\kappa$
between the two subsystems~\cite{FRV-22}.  This scenario represents
a useful playground where to study the mutual interplay and the
properties of different subparts of a complex quantum many-body
system, such as the onset of decoherence or the exchange of heat and
work between them.  In particular, the set of $n$ qubits can be
interpreted as a probing apparatus ($\mathcal{A}$) for the set of $L$
spins composing the Ising ring ($\mathcal{I}$).

By using proper dynamic FSS frameworks~\cite{RV-21} we show that, at
the Ising CQT ($g = g_{\mathcal I}$) or along its FOQT line
($|g| < g_{\mathcal I}$), peculiar dynamic scaling regimes may appear.
To this purpose, we explicitly consider two different large-size limits
where dynamic FSS laws emerge: we allow $L\to\infty$ while fixing either
the number $n$ of external qubits, or their nearest-neighbor interspace
distance $b$.  For a sufficiently large number $n$ of external spins,
the dependence of observables on the number $n$ of qubits can be
reabsorbed into a redefinition of the quench parameter, by replacing
the scaling variable associated with $\kappa$ with $\sqrt{n} \,
\kappa$.

We also consider quenches within the Ising-disordered phase ($g>
g_{\mathcal I}$), addressing the role of the $\sqrt{n}$-behavior out
of the FSS framework. We observe that, by maintaining $\kappa\sqrt{n}$
fixed, which represents the typical energy scale of work done on the
whole system [in particular $\sqrt{\expval{W^2}}=\kappa\sqrt{n}$,
  cf. Eq.~(\ref{workav})], several quantities that are expected to
scale extensively with $n$ approach a constant value in the large
$L$ limit.  In other words, our results suggest that even beyond the
FSS regime, there is a fair tradeoff between the {\em linear}
enlargement of both the Ising ring and ancillary spins (they both
increase with $L$ when $b$ is fixed) and the $\sqrt{n}$-reduction of
the quench intensity.

Finally, we compare our results to those obtained when also a
nearest-neighbor coupling among the external qubits is taken into
account, maintaining the Ising-ring length fixed.  For quenches with
$\kappa:0\to1$, no substantial differences emerge between the two
lattice models: although the recharging of the {\em coupled} model is
faster than that of the {\em free} system, the two models can roughly
store the same amount of energy. However, our study was only
exploratory, we have not done a through optimization study of the
model parameters.

Our dynamic framework has been carefully tested by means of numerical
exact diagonalization techniques on systems with only $\sim O(20)$
quantum spins.  Moreover, results for the time behavior of the
decoherence function are strictly connected with the experimentally
measurable R{\'e}nyi entropy~\cite{Islam-etal-2015,
  Brydges-etal-2019}, while the average energy stored into the
external qubits may represent a central thermodynamic quantity to be
monitored in the study of quantum-battery devices~\cite{CPV-18}.
The reduced number of spins in our simulated systems suggests that
quantum simulators with existing technology should be able to access
our predictions, e.g., by using trapped ions~\cite{Debnath-etal-16},
ultracold atoms~\cite{Labuhn-etal-16, SFSTT-20}, or superconducting
qubits~\cite{Salathe-etal-15, Cervera-18}.

%%%%%%%%%%%%
\section*{References} %
%%%%%%%%%%%%

\end{document}